\documentclass[%
reprint,
superscriptaddress,
 amsmath,amssymb,
 aps,
prx,
]{revtex4-2}

\usepackage{graphicx}
\usepackage{dcolumn}
\usepackage{bm}
\usepackage{dblfloatfix}
\usepackage{gensymb}
\usepackage{xcolor}
\usepackage{amsmath}
\usepackage{amssymb}
\usepackage{units}
\usepackage[english]{babel}

\begin{document}

\title{Controlled introduction of defects to delafossite metals by electron irradiation}

\author{V. Sunko}
\thanks{V. S. and P. H. M. contributed equally to this work.}
\email{Veronika.Sunko@cpfs.mpg.de}
\affiliation {Max Planck Institute for Chemical Physics of Solids, N{\"o}thnitzer Stra{\ss}e 40, 01187 Dresden, Germany}
\affiliation {SUPA, School of Physics and Astronomy, University of St. Andrews, St. Andrews KY16 9SS, United Kingdom}

\author{P. H. McGuinness}
\thanks{V. S. and P. H. M. contributed equally to this work.}
\email{Philippa.McGuinness@cpfs.mpg.de}
\affiliation {Max Planck Institute for Chemical Physics of Solids, N{\"o}thnitzer Stra{\ss}e 40, 01187 Dresden, Germany}
\affiliation {SUPA, School of Physics and Astronomy, University of St. Andrews, St. Andrews KY16 9SS, United Kingdom}

\author{C. S. Chang}
\affiliation {School of Applied and Engineering Physics, Cornell University, Ithaca, New York 14853, United States}
\affiliation {Department of Physics, Cornell University, Ithaca, New York 14853, United States}

\author{E. Zhakina}
\affiliation {Max Planck Institute for Chemical Physics of Solids, N{\"o}thnitzer Stra{\ss}e 40, 01187 Dresden, Germany}
\affiliation {SUPA, School of Physics and Astronomy, University of St. Andrews, St. Andrews KY16 9SS, United Kingdom}

\author{S. Khim}
\affiliation {Max Planck Institute for Chemical Physics of Solids, N{\"o}thnitzer Stra{\ss}e 40, 01187 Dresden, Germany}

\author{C. E. Dreyer}
\affiliation {Department of Physics and Astronomy, Stony Brook University, Stony Brook, New York 11794-3800, United States}
\affiliation {Center for Computational Quantum Physics, Flatiron Institute, 162 5th Avenue, New York, New York 10010, United States}

\author{M. Konczykowski}
\affiliation {Laboratoire des Solides Irradiés, CEA/DRF/lRAMIS, Ecole Polytechnique,CNRS, Institut Polytechnique de Paris, F-91128 Palaiseau, France}

\author{M. K\"{o}nig}
\affiliation {Max Planck Institute for Chemical Physics of Solids, N{\"o}thnitzer Stra{\ss}e 40, 01187 Dresden, Germany}

\author{D. A. Muller}
\affiliation {School of Applied and Engineering Physics, Cornell University, Ithaca, New York 14853, United States}
\affiliation {Kavli Institute at Cornell for Nanoscale Science, Ithaca, New York 14853, United States}

\author{A. P. Mackenzie}
\email{Andy.Mackenzie@cpfs.mpg.de}
\affiliation {Max Planck Institute for Chemical Physics of Solids, N{\"o}thnitzer Stra{\ss}e 40, 01187 Dresden, Germany}
\affiliation {SUPA, School of Physics and Astronomy, University of St. Andrews, St. Andrews KY16 9SS, United Kingdom}

\begin{abstract}

The delafossite metals PdCoO$_{2}$, PtCoO$_{2}$ and PdCrO$_{2}$ are among the highest conductivity materials known, with low temperature mean free paths of tens of microns in the best as-grown single crystals.  A key question is whether these very low resistive scattering rates result from strongly suppressed backscattering due to special features of the electronic structure, or are a consequence of highly unusual levels of crystalline perfection.  We report the results of experiments in which high energy electron irradiation was used to introduce point disorder to the Pd and Pt layers in which the conduction occurs.  We obtain the cross-section for formation of Frenkel pairs in absolute units, and cross-check our analysis with first principles calculations of the relevant atomic displacement energies.  We observe an increase of resistivity that is linear in defect density with a slope consistent with scattering in the unitary limit.  Our results enable us to deduce that the as-grown crystals contain extremely low levels of in-plane defects of approximately $0.001\%$.  This confirms that crystalline perfection is the most important factor in realizing the long mean free paths, and highlights how unusual these delafossite metals are in comparison with the vast majority of other multi-component oxides and alloys. We discuss the implications of our findings for future materials research.
\end{abstract}

\maketitle

\section{\label{sec:Intro}Introduction}

Throughout the early evolution of the physics of metals, it was assumed that extremely high metallic state electrical conductivity would be restricted to elemental metals, which could be purified and then annealed to remove dislocations and other structural defects, resulting in long electron mean free paths.  After decades of materials research and refinement, techniques such as heterodoping of semiconductor devices led to the creation of extremely long mean free paths in two-dimensional electron gases (2DEGs), most notably those fabricated at the GaAs/GaAlAs interface.  Although painstaking refinement of semiconductor 2DEGs has continued, resulting in exquisite levels of purity \cite{gardner_modified_2016}, the past decade has seen rapid parallel developments in other materials.  The insights obtained from the investigations of graphene \cite{banszerus_ballistic_2016} and of topological insulators \cite{hasan_topological_2010, qi_topological_2011} led to rapid advances in the study of Dirac and Weyl semimetals \cite{armitage_weyl_2018}, in which resistivities in the n$\Omega$cm range have been observed \cite{liang_ultrahigh_2015, shekhar_extremely_2015, kumar_extremely_2017}. These spectacular observations, however, result from materials physics different from that of normal semiconductors.  Rather than being materials of particularly high perfection, the main reason for the low resistivity of Weyl and Dirac semi-metals is protection from backscattering due to the exotic features of both their bulk band structure and the surface states that they host \cite{zhang_ultrahigh_2019}. Strong evidence in favour of this interpretation comes from the ratio of the resistive mean free path to that determined by analysis of de Haas van Alphen (dHvA) oscillations.  The resistive mean free path, which is mainly sensitive to backscattering, is a factor of $\sim10^4$ larger than the dHvA-derived one, which is limited by small-angle scattering events that make only a small contribution to resistivity \cite{liang_ultrahigh_2015, shekhar_extremely_2015, kumar_extremely_2017}.

A further class of materials recently discovered to have extremely long mean free paths at low temperature are the delafossite oxide metals PdCoO$_{2}$, PtCoO$_{2}$ and PdCrO$_{2}$ \cite{shannon_chemistry_1971, mackenzie_properties_2017}.  Each can have resistive mean free paths of microns at low temperatures \cite{takatsu_roles_2007, takatsu_anisotropy_2010, kushwaha_nearly_2015}.  Indeed, in the most conductive single crystals of PdCoO$_{2}$, with resistivities as low as $\unit[8]{n\Omega cm}$, mean free paths as long as $\unit[20]{\mu m}$ have been reported \cite{hicks_quantum_2012}.  These extremely long mean free paths have led to the observation of a number of novel transport properties \cite{takatsu_extremely_2013, daou_large_2015, moll_evidence_2016, kikugawa_interplanar_2016, nandi_unconventional_2018, putzke_h/e_2019} and there is every prospect of further discoveries.  The observation that metallic oxides could have mean free paths as long as this is surprising, especially since they appear in crystals grown using fairly standard methods and not subject to any post-growth annealing.  Put at its simplest, the key question is whether the huge resistive mean free paths are primarily the result of scattering suppression, as in Dirac and Weyl materials, or of unprecedented levels of crystalline perfection.  A first clue to the answer to this question came from a study of scattering suppression due to momentum-orbital locking around the Fermi surface of PtCoO$_{2}$ \cite{usui_hidden_2019}. Scattering is predicted to be suppressed by a factor of order 2-4 from that in a material with a trivial Fermi surface of the same size.  Although this aspect of the physics of the delafossite metals is important, the predicted scattering suppression is nowhere near sufficient to account for the observed low temperature mean free paths. Consistent with this finding, the ratio between the resistive and dHvA mean free paths is a much more modest factor of 10-20 \cite{hicks_quantum_2012, kushwaha_nearly_2015} than the $10^4$ reported in the Dirac and Weyl materials. In combination, the transport data and the analysis presented in Ref.\ \cite{usui_hidden_2019} suggest that the delafossite metals might naturally exist with a level of purity that is more or less never observed in as-grown crystals.  If that were the case, it would be extremely surprising, and open new avenues of research in quantum materials.  

In this paper, we report on a combined experimental and theoretical investigation of the disorder dependence of resistivity in PtCoO$_{2}$, PdCoO$_{2}$ and PdCrO$_{2}$. High resolution scanning transmission electron microscopy (STEM) confirms the long-range structural purity of single crystals, which contain very few layer defects or dislocations.  No point defects such as intersite substitutions or vacancies could be resolved by STEM, but the detection sensitivity for such defects is not sufficient to fully address the issue of ultra-high purity.  To investigate that in depth, we turn to deliberate irradiation.  Using $\unit[2.5]{MeV}$ electrons, we create controlled densities of ‘Frenkel pairs’, i.e. displaced atoms leading to a vacancy and an interstitial atom. The kinetics of our experiment are such that no damage cascades are created, and genuine point disorder is achieved.  By calculating the absolute cross-section for Frenkel pair creation, we prove that the defect scattering is not hugely suppressed in these metals, and show that the point defect density in the as-grown crystals is approximately $0.001\%$.  In other words, the extremely long mean free paths of the delafossite metals are indeed primarily due to a highly unusual level of crystalline perfection.  In order to investigate this remarkable fact, and cross-check the analysis that was used to deduce it, we perform first-principles calculations of the atomic displacement energies in delafossites.

The paper is organized as follows.  In section \ref{sec:TEM} we describe the delafossite crystal structure and sample growth, and show transmission electron micrographs of as-grown crystals.  We then describe the irradiation experiments and the calculation of the Frenkel pair production cross-section in section \ref{sec:ElIrr}, before discussing the dependence of resistivity on defect density in section \ref{sec:REsistivityDefect}.  First-principles calculations are presented in section \ref{sec:FirstPrinciples}, before we close the paper with examples of the changes to transport properties produced by the irradiation in section \ref{sec:Transport}, a discussion and conclusions in sections \ref{sec:Discussion} and \ref{sec:Conclusions}, respectively.

\section{\label{sec:TEM} Sample growth and imaging}

\begin{figure}[h]
\includegraphics{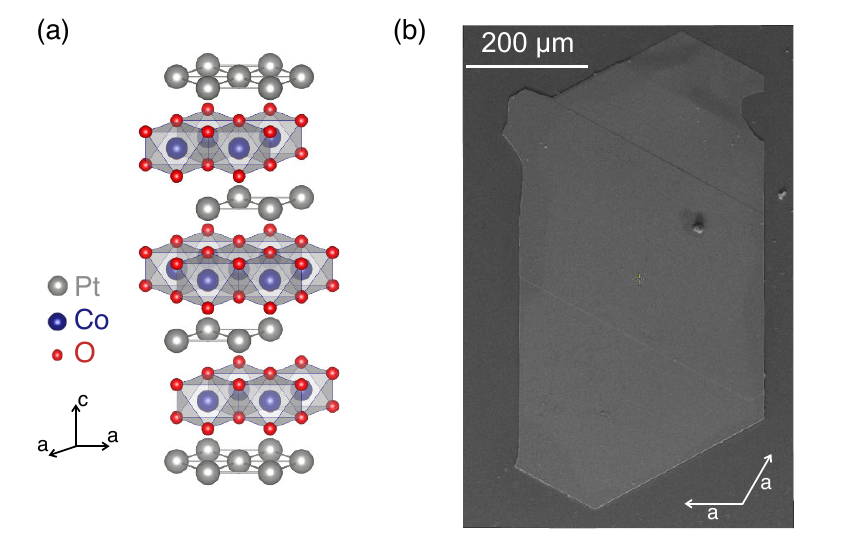}
\caption{\label{fig:SEM} (a) Delafossite crystal structure, shown on the example of PtCoO$_{2}$. (b) Scanning electron microscopy (SEM) image of a crystal of PtCoO$_{2}$.}
\end{figure}

The delafossite crystal structure is shown in Fig.\ \ref{fig:SEM}a.  It is highly anisotropic, with the Pt or Pd layers separated by layers of cobalt oxide.  Each layer is triangularly co-ordinated, and the interlayer repeat gives the space group $R\overline{3}m$ ($D_{3d}^5$).  Highly two-dimensional conduction ($\rho_{c}/\rho_{ab}\sim10^3$) takes place in the Pd/Pt layers, while the CoO$_2$ layers can in the first approximation be thought of as insulating.  It is clear from these considerations that point disorder in the Pd/Pt layers is likely to dominate the scattering, a point to which we return below.

Single crystals of PtCoO$_{2}$, PdCoO$_{2}$ and PdCrO$_{2}$ were grown in sealed quartz tubes using methods discussed in \cite{kushwaha_nearly_2015, kushwaha_single_2017}. The crystals grow as platelets of $\sim \unit[1-10]{\mu m} $ in thickness, and $\sim \unit[300-700]{\mu m}$ in lateral dimensions. The crystal edges typically close a 120$\degree$ angle with each other, as shown in Fig.\ \ref{fig:SEM}b, and are perpendicular to the crystalline axes. Step edges between terraces of uniform thickness are clearly visible, enabling us to perform in-plane transport measurements on a single terrace, therefore avoiding contributions from the much higher $c$-axis resistivity. 

\begin{figure}[t]
\includegraphics{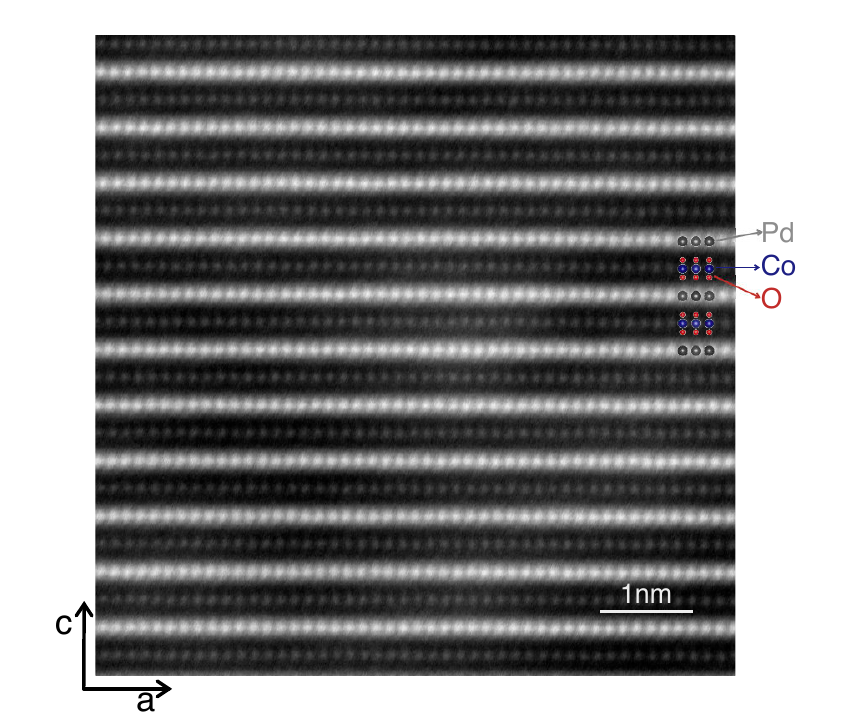}
\caption{\label{fig:TEM} A STEM image of a pristine PdCoO$_{2}$ sample taken at high magnification resolving Pd and Co atomic sites using High Angle Annular Dark Field (HAADF-STEM) imaging.}
\end{figure}

Such samples can be thinned down using a focused ion beam (FIB) with standard procedures in order to image them using Scanning Transmission Electron Microscopy. In Fig.\ \ref{fig:TEM} we show a high angle annular dark field STEM image of a PdCoO$_{2}$ sample, in which atomic contrast has been achieved. No point defects are visible in the image, which is representative of multiple measurements on several samples. This of course does not rule out the existence of point defects, such as interstitials or vacancies, but it does indicate that the concentration of such defects is beyond the resolution of STEM. 

\section{\label{sec:ElIrr} Electron irradiation}

Because the defect concentration in as-grown crystals is too low to be directly determined, investigating its influence on resistivity requires deliberate and controlled introduction of disorder. Crucially, in order to mimic the situation in the as-grown materials, no large voids or columnar defects should be created, nor should foreign atoms be implanted in the crystal. Instead, individual point defects should be introduced. Irradiation by high-energy electrons in the energy range of $\unit[1-10]{MeV}$ is the ideal technique to achieve this type of disorder: electrons of these energies can transmit enough energy to an atom to displace it from its lattice site, but not enough for the displaced atom to create a significant number of additional defects. Therefore, individual Frenkel pairs are created. This is in contrast to ion irradiation, where the larger mass of the incoming particle turns every collision into a collision cascade, creating large columnar defects.

\subsection{\label{sec:elIrr_technique} Experimental set-up}

\subsubsection{\label{sec:Beamline} Beamline}
\begin{figure}[h]
\includegraphics{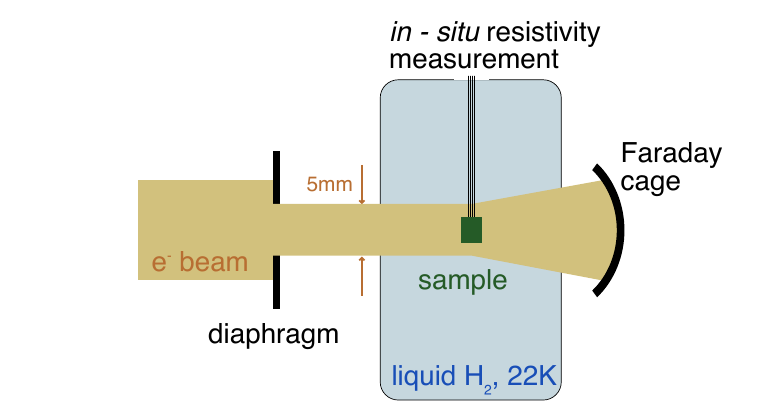}
\caption{\label{fig:Beamline} A sketch of the experimental set-up for electron irradiation. The yellow area indicates the range in which the electron beam is swept, rather than representing the beam cross-section.}
\end{figure}

The irradiation with electrons with a maximum kinetic energy of $\unit[2.5]{MeV}$ was performed at the SIRIUS Pelletron linear accelerator operated by the Laboratoire des Solides Irradi\'{e}s (LSI) at the Ecole Polytechnique in Palaiseau, France. A sketch of the experimental set-up is shown in Fig.\ \ref{fig:Beamline}. During the irradiation the sample was immersed in a bath of liquid hydrogen at a temperature of $ \approx \unit[22]{K}$, ensuring that the introduced defects are not mobile. The accelerator beam carried a current in the range of $\unit[5 - 8]{\mu A}$, and passed through a circular diaphragm aperture of a $\unit[5]{mm}$ diameter before reaching the sample. The current passing through the diaphragm was experimentally determined using a control metallic sample, and found to be in the range of $\unit[1.5 - 2.5]{\mu A}$, corresponding to current densities at the sample location of $\unit[8 -13]{\mu A/cm^2}$.  The beam was swept vertically and horizontally at two incommensurate frequencies, ensuring the homogeneous irradiation of the sample area. The homogeneous irradiation throughout the sample thickness was guaranteed by the large  penetration range of electrons, estimated to be $\sim \unit[1.8]{mm}$ for $\unit[2.5]{MeV}$ electrons in PtCoO$_{2}$ \cite{berger_star_2019}, and thus two to three orders of magnitude larger than the typical thickness of our crystals.  This further ensures that the majority of  electrons can be measured using a Faraday cage placed behind the sample stage, enabling the monitoring of the current fluctuations during a measurement (Fig.\ \ref{fig:Beamline}). 

The irradiation is paused at regular intervals to perform four point \emph{in-situ} resistance measurements, and therefore monitor the increase of resistivity as a function of electron dose. Measuring resistivity of crystals as conductive and as small as delafossites in those demanding conditions presents additional challenges, which we addressed by developing dedicated sample preparation methods, as described below. 

\subsubsection{\label{sec:FiB} Sample preparation}

\begin{figure}[h]
\includegraphics{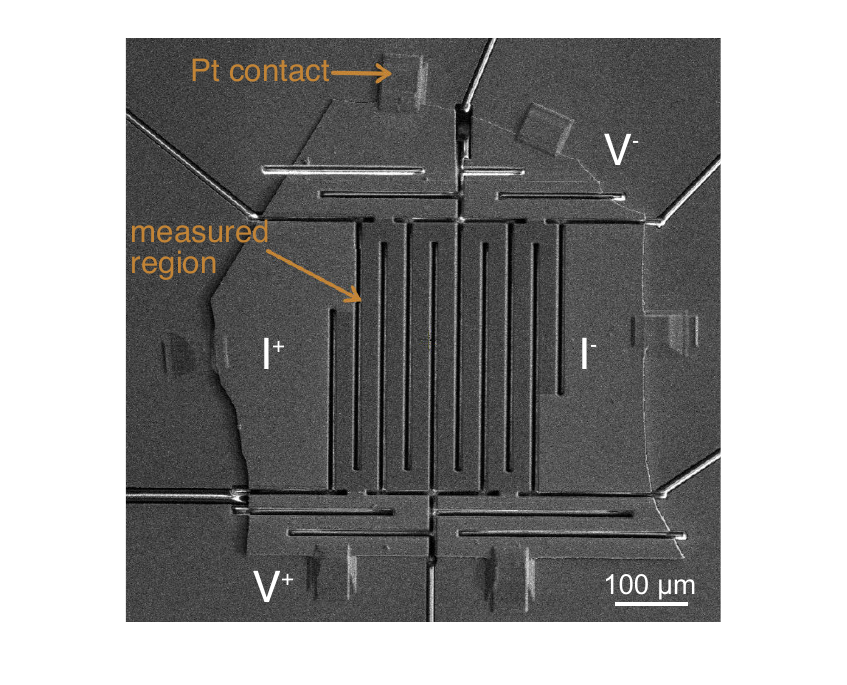}
\caption{\label{fig:FIBsample} A scanning electron microscopy (SEM) image of a microstructured PtCoO$_{2}$ sample used for the \emph{in-situ}  measurements of resistivity. The measured region of the device appears darker in the SEM image because the gold has been removed from it; the rest of the sample, the Pt contacts and the substrate are covered with sputtered gold.}
\end{figure}

In order to enable a reliable \emph{in-situ} measurement of the low-temperature resistivity of delafossites, we used focused ion beam sculpting to increase the effective length of the measured sample, as shown in Fig.\ \ref{fig:FIBsample}. The standard method of sample mounting for FIB sculpting requires using a layer of glue to attach the crystal to a substrate, as was done in previous transport studies of delafossite metals \cite{moll_evidence_2016, nandi_unconventional_2018}. However, glue degrades quickly in the electron beam, motivating us to use a glue-free mounting method instead. The sample, held on a $\sim \unit[25]{\mu m}$ thick mica substrate by the electrostatic force, was first covered with a $\unit[150]{nm}$ thick layer of sputtered gold. Pt contacts were then deposited \emph{in-situ} in the FIB, providing a mechanical connection between the sample and the gold-covered substrate. To achieve  contact resistances on the order of $\unit[1]{\Omega}$ a second layer of gold was sputtered over the sample, substrate and the Pt contacts.  A crystal mounted in this way was FIB-structured using the standard techniques described in detail in Ref.\ \onlinecite{moll_focused_2018}, including the removal of gold from the measured part of the device. This active region is shaped like a meander, whose width of $\unit[60]{\mu m}$ was chosen to ensure that the edge scattering does not significantly contribute to the measured resistivity \cite{moll_evidence_2016}. The length of the meander is approximately $\unit[3]{mm}$ in the example structure shown in Fig.\ \ref{fig:FIBsample}. The  two current contacts (labelled $I^-$ and $I^+$ in Fig.\ \ref{fig:FIBsample}) and two of the voltage contacts ($V^-$ and $V^+$) were used for the \emph{in-situ} measurement. Additional meanders were cut in all the voltage contacts so that the strain caused by the differential thermal contraction between the sample and the substrate can relax. Consequently, the samples were mechanically stable enough to allow for multiple iterations of irradiation and rapid thermal cycling. The PtCoO$_{2}$ samples prepared in this way had low temperature resistances in the range of  $\unit[1 - 20]{m\Omega}$.

\subsection{\label{sec:ResIncrease} Resistivity increase}

\begin{figure}[h]
\includegraphics[scale=.9]{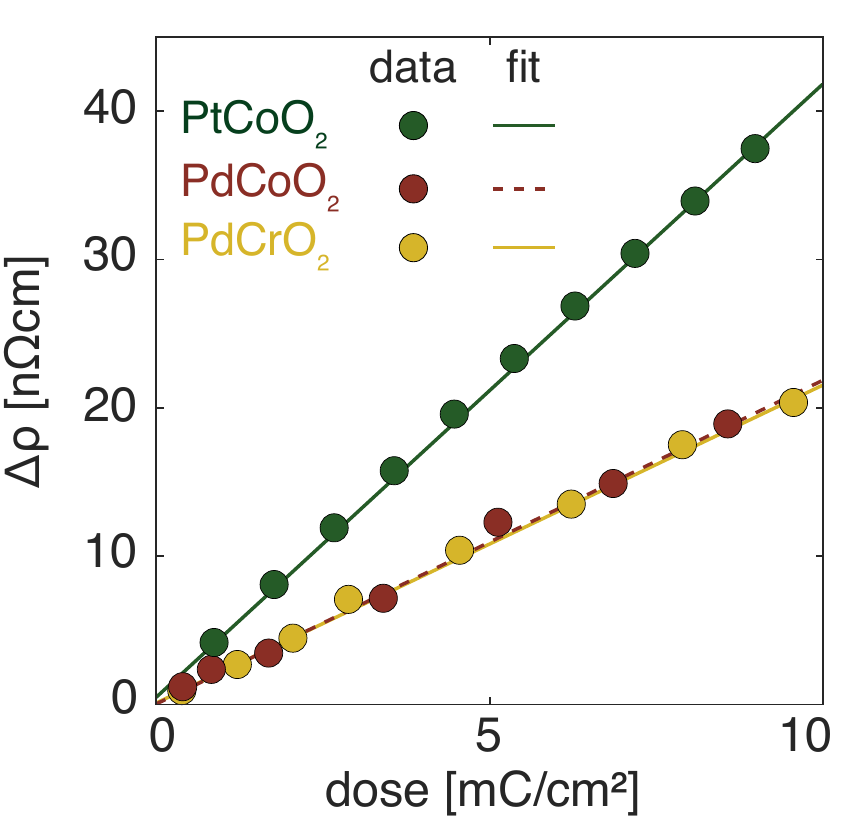}
\caption{\label{fig:ResisitivityIncrease} Increase of resistivity as a function of electron dose, for three delafossite metals: PtCoO$_{2}$, PdCoO$_{2}$ and PdCrO$_{2}$.}
\end{figure}

The increase of the resistivity of PtCoO$_{2}$, PdCoO$_{2}$ and PdCrO$_{2}$ as a function of electron dose is shown in Fig.\ \ref{fig:ResisitivityIncrease}. In all three compounds the dependence of resistivity on dose is linear in the investigated range. The rate of resistivity increase is higher in PtCoO$_{2}$ than in PdCoO$_{2}$ and PdCrO$_{2}$, which show the same rate of increase. These observations indicate that the resistivity is dominated by the defects in the conductive Pt/Pd planes, as expected in such two-dimensional systems: Pd defects are created at the same rate in PdCoO$_{2}$ and PdCrO$_{2}$, while the rate of defect introduction is higher in PtCoO$_{2}$, because the larger nuclear charge of the Pt atoms leads to a stronger interaction with the incoming electrons. The higher rate of defect introduction is the reason we have chosen to perform further measurements on PtCoO$_{2}$. 

The fact that the resistivity increases as a function of electron dose indicates both that electron irradiation introduces defects, and that those defects contribute to resistivity. Therefore, if the scattering of conduction electrons by defects is suppressed in delafossites, that suppression is not complete. However, the observations so far do not rule out a partial suppression of scattering, and therefore cannot be used to deduce how sensitive the resistivity of delafossites is to disorder. This requires quantifying the introduced defect concentration, which we have done using electron-energy dependent measurements, as described in the following section.  

\subsection{\label{sec:Quant} Quantifying the introduced Frenkel pair concentration}

In order to calculate the concentration of introduced Frenkel pairs we adopt the approach used in the studies of electron irradiation of elemental metals and their alloys \cite{iseler_production_1966}, as well as in the high-temperature superconductor YBCO \cite{legris_effects_1993}, and graphene \cite{meyer_accurate_2012}. The method relies on the  electron-energy dependence of the cross-section for the formation of Frenkel pairs, $\sigma_{Fp}$. A quantity directly proportional to $\sigma_{Fp}$, such as the rate of increase of resistivity, is measured as a function of electron energy. This measurement is compared to a calculation of the electron-energy dependence of $\sigma_{Fp}$, revealing the factor relating the measured quantity and $\sigma_{Fp}$, therefore allowing the determination of the latter in absolute units. 

This approach relies on the calculation of $\sigma_{Fp}$, which requires the knowledge of the cross-section for the scattering of electrons off stationary nuclei, and understanding of the conditions under which such a scattering event results in an atom being displaced from its equilibrium lattice site. If the incoming particles are non-relativistic, the former is given by the well-known Rutherford expression for a differential scattering cross-section:

\begin{equation}
\label{ruth}
\text{\ensuremath{\frac{d\sigma_{R}}{d\Omega}}}=\left(\frac{Ze^{2}}{4\pi\varepsilon_{0}}\frac{\hbar c}{4E_{K}}\right)^{2}\frac{1}{\sin^{4}\left(\frac{\vartheta}{2}\right)},
\end{equation}
where $E_{K}$ is the electron kinetic energy, $\vartheta$ the scattering angle, $Z$  the atomic number of the nucleus, $e$ the elementary charge, $\varepsilon_{0}$ the vacuum permeability, $\hbar$ the reduced Planck's constant and $c$ the speed of light. 

However, the high energy of the electrons in the present experiment requires the use of the relativistic extension of the Rutherford expression. Mott calculated the relativistic differential cross-section for the scattering of electrons and point-like nuclei, using Darwin's solution to the Dirac equation \cite{mott_scattering_1929, mott_polarisation_1932}. His formula is exact, and it gives the cross-section as a sum of two conditionally convergent infinite series. It therefore needs to be numerically evaluated for every individual case, motivating numerous approximate expressions for the Mott cross-section, such as the commonly used McKinley-Feshbach cross-section \cite{mckinley_coulomb_1948}. However, the McKinley-Feshbach approximation is only valid for atoms of the nuclear charge $Z<27$. Although it is often used for heavier atoms, such as copper \cite{iseler_production_1966}, it is not expected to be suitable for Pd or Pt. Instead, we follow the approach given in Refs. \onlinecite{lijian_analytic_1995, boschini_expression_2013}, in which the ratio of the Mott and Rutherford cross-section is expressed as a function of scattering angle ($\vartheta$) and the ratio of the electron velocity to the speed of light $\beta$ \footnote{$\beta=\sqrt{1-(m_{e}c^2/(m_{e}c^2+E_{K}))^2}$, where $m_{e}$ is the electron rest mass. } as: 

\begin{equation}
\label{cross}
\frac{\sigma_{M}}{\sigma_{R}}={\sum\limits_{j=0}^4}{\sum\limits_{k=1}^6}b_{jk}\left(\beta-\overline{\beta}\right)^{k-1}\left(1-\cos\left(\vartheta\right)\right)^{j/2},
\end{equation}
where $\overline{\beta}=0.7181287$. $b_{jk}$ are a set of thirty parameters, obtained by fits to the numerical solutions to  Mott's equations. They are listed in Ref. \onlinecite{lijian_analytic_1995} for all elements with $Z\leq90$ and in Ref. \onlinecite{boschini_expression_2013} for all elements with $Z\leq118$; for more details on the mathematics and numerics behind those calculations we refer the reader to Refs. \onlinecite{lijian_analytic_1995, boschini_expression_2013} and the references within. In this work we calculate the Mott differential cross-section using the expressions \ref{ruth} and \ref{cross}, combined with parameters $b_{jk}$ as listed in Ref. \onlinecite{boschini_expression_2013}. 

A scattering event described by the Mott cross-section results in a Frenkel pair if, and only if, the energy transferred to the nucleus is larger than a threshold, the  so-called the displacement energy  ($E_{d}$). As long as the displaced atom does not introduce secondary defects, the total cross-section for production of Frenkel pairs is given by: 

\begin{equation}
\label{FP}
\sigma_{Fp}=2\pi\int\limits_{\vartheta(E_{D})}^{\pi}{\frac{d\sigma_{M}}{d\Omega}\sin\vartheta d\vartheta},
\end{equation}
where $\vartheta(E_{D})$ denotes the scattering angle corresponding to the minimal energy transfer of $E_{d}$ \footnote{The scattering angle $\vartheta$ and the transferred energy $E$ are related by: $E=E_{max}\sin^2{\vartheta/2}$. $E_{max}$ is the maximum energy that can be transferred, $E_{max}=2E_{K}\left(E_{K}+2m_{e}c^2\right)/Mc^2$. $E_{K}$ is the kinetic energy of the incoming electron, and $m_{e}$ and $M$ the masses of the electron and the nucleus, respectively.}.

\begin{figure}[h]
\includegraphics[scale=0.9]{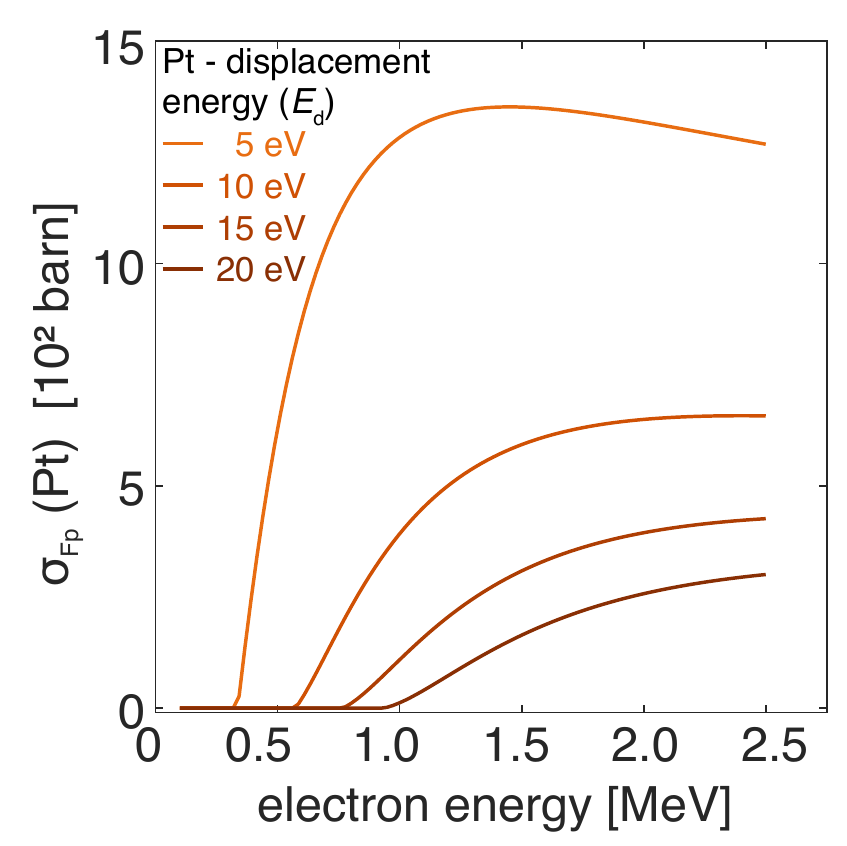}
\caption{\label{fig:PtDispl} Frenkel-pair production cross-section for Pt atoms as a function of electron energy, for a range of displacement energies.}
\end{figure}

The calculation described above has only one free parameter, the displacement energy. In Fig.\ \ref{fig:PtDispl} we show the  cross-section for production of Pt Frenkel pairs calculated for a range of displacement energies. For each of them there is a \emph{different} well-defined minimum electron energy at which Frenkel pairs can be created. Consequently, the curves at different values of $E_{d}$ are not related by a simple scaling relation, and the displacement energy can be uniquely determined by a measurement of the electron-energy dependence of any quantity which is proportional to the Frenkel pair production cross-section.  

\begin{figure}[h]
\includegraphics[scale=1]{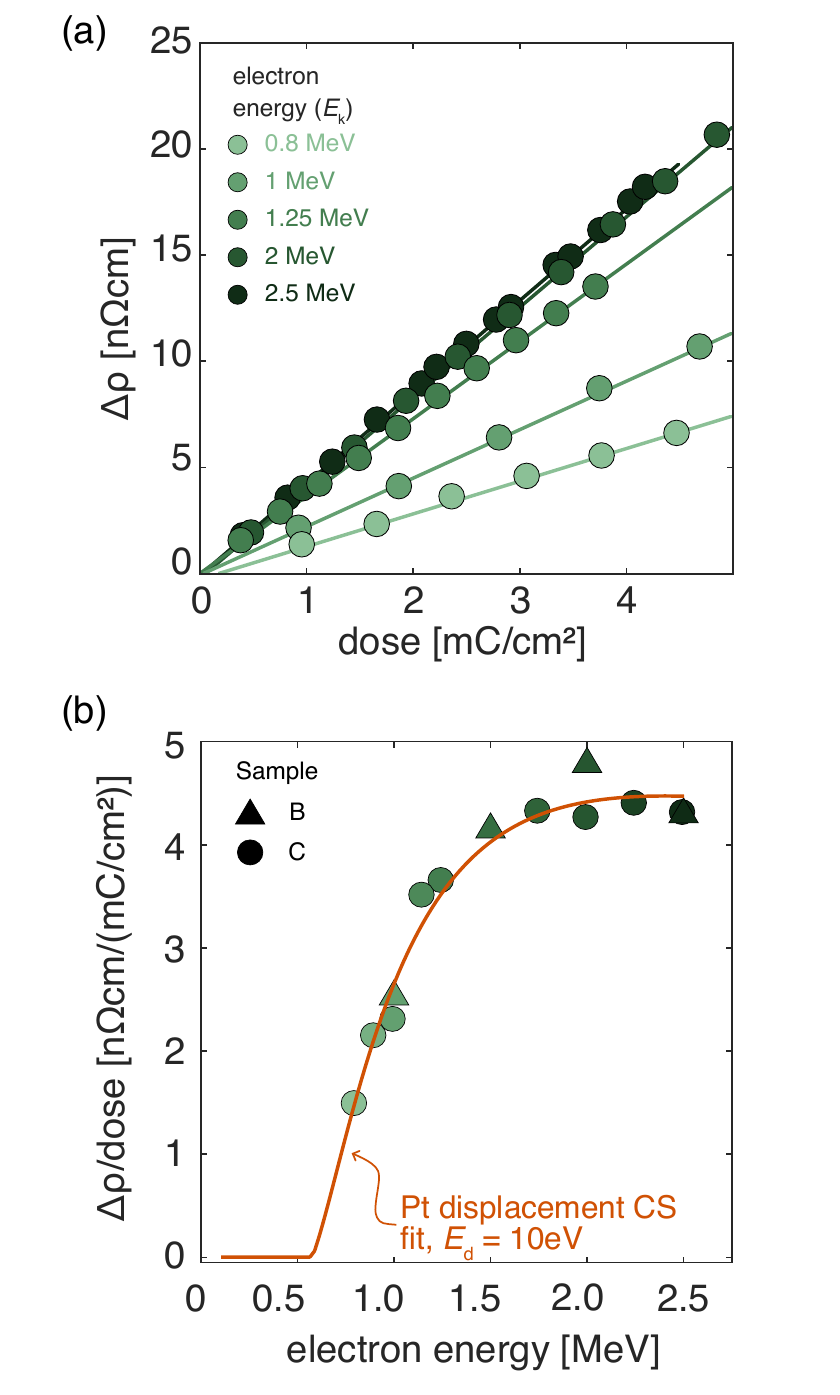}
\caption{\label{fig:ElEnergy} (a) PtCoO$_{2}$ resistivity as a function of electron dose for a range of electron energies. (b) The rate of increase of the resistivity of PtCoO$_{2}$ as a function of electron energy, measured on two samples (symbols), compared to the calculated cross-section for the displacement energy of $10\unit{eV}$.}
\end{figure}

The rate of resistivity increase as a function of dose is such a quantity. We have therefore irradiated two samples of PtCoO$_{2}$ using a range of electron energies between $0.8$ and $\unit[2.5]{MeV}$ \footnote{technically the accelerator energy can be reduced down to $\unit[0.3]{MeV}$, however this requires a stabilisation time of a few days, and was therefore not feasible in the limited beamtime}. As shown in Fig.\ \ref{fig:ElEnergy}a, the resistivity increases more slowly when the electron energy is decreased, as expected. The slopes measured on the two samples are plotted as a function of electron energy in Fig.\ \ref{fig:ElEnergy}b (symbols). They were fitted to the Pt Frenkel pair production cross-section, as calculated by the methods described above, determining the displacement energy to be equal to $E_{d}=\unit[10]{eV}$. The good agreement of the data and the calculation both justifies the assumption that significant numbers of secondary defects are not created and confirms that the resistivity is dominated by Pt defects. What is more, it allows for a determination of  the Frenkel pair production cross-section in absolute units; for electrons of kinetic energy of $\unit[2.5]{MeV}$ it is $\sigma_{Fp}\left(\text{Pt}, \unit[2.5]{MeV}\right)=\unit[657]{barn}$. Multiplying the irradiation dose by $\sigma_{Fp}$, and dividing it by the charge of an electron, directly yields the introduced Frenkel pair concentration. 

\section{\label{sec:REsistivityDefect} Resistivity as a function of Frenkel pair concentration}
\begin{figure}[b]
\includegraphics[scale=1]{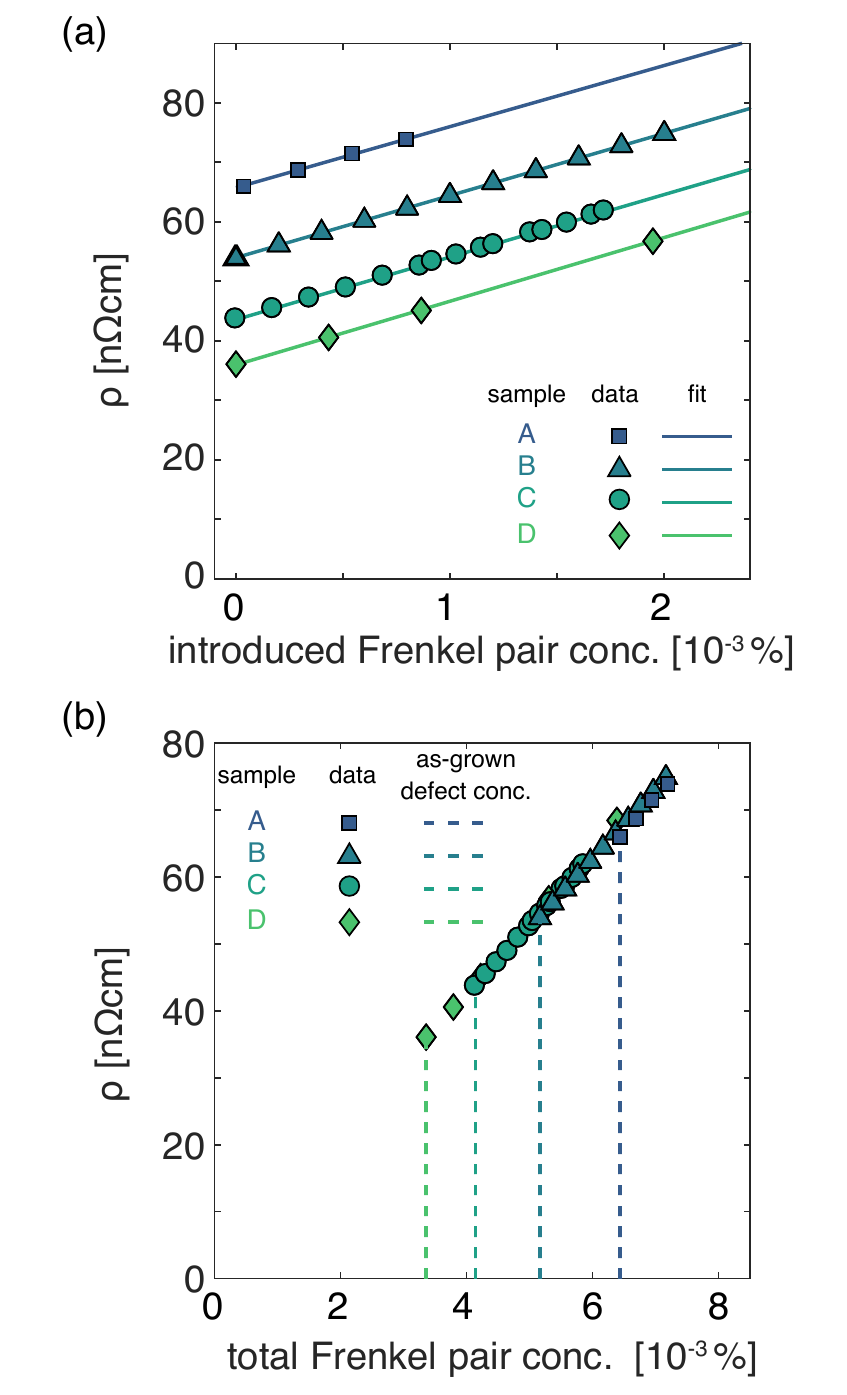}
\caption{\label{fig:ResDef} Resistivity of four PtCoO$_{2}$ samples as a function of (a)  introduced defect concentration, and (b) the total defect concentration, estimated from the interpolation of lines in (a). The dashed lines indicate the effective Frenkel pair concentration in the samples before irradiation. }
\end{figure}

In Fig.\ \ref{fig:ResDef}a we show the dependence of resistivity on the concentration of Frenkel pairs introduced by electron irradiation for four PtCoO$_{2}$ samples microstructured into long meanders, as shown in Fig.\ \ref{fig:FIBsample}. The dependence is linear in all the samples, with a slope of $(9.1\pm0.2)\times 10^3\unit{n\Omega cm/\%}$. The residual resistivity of the samples prior to irradiation, however, varied between $40$ and $\unit[65]{n\Omega cm}$, reflecting different defect concentration in as-grown crystals. Assuming that all the defects that contribute to resistivity are Frenkel pairs, our measurements allow us to estimate their concentration prior to irradiation  by extrapolating the lines in Fig.\ \ref{fig:ResDef}a to zero resistivity. Once plotted as a function of total Frenkel pair concentration, found by adding the initial and the introduced defect concentrations, the resistivities of all the samples collapse on the same curve, as shown in  Fig.\ \ref{fig:ResDef}b. 
The estimated initial Frenkel pair concentrations, indicated by the dashed lines, range between $0.004$ and $0.007\%$ in the investigated samples, while the lowest low-temperature resistivity reported in PtCoO$_{2}$ ($20\unit{n\Omega cm}$) \cite{nandi_unconventional_2018} corresponds to a defect concentration of $0.002\%$. The assumption that all the defects contributing to resistivity are Frenkel pairs is certainly not entirely accurate, but other point defects in the Pt layers are not expected to have a significantly different influence on resistivity. The estimate for the initial point defect concentration obtained in this way should therefore be correct to within a factor of 2-3. 

\subsection{\label{sec:OtherMaterials} Comparison with other 2D materials}
In addition to enabling the determination of the purity of as-grown crystals, the analysis presented in the previous section allows the sensitivity of the delafossite resistivity to point defects to be compared to that of other materials. The sheet resistivity $\rho^{2D}$ of 2D materials is usually assumed to depend on the ratio of the in-plane defect concentration $n_{d}$ and carrier concentration $n$ as: 
\begin{equation}\label{eq:unitary}
\rho_{unit}^{2D}=\frac{4\hbar}{e^{2}}\frac{n_{d}}{n}, 
\end{equation}
where $e$ is the electron charge. This expression is referred to as the `unitary limit', and physically corresponds to the case of the strongest possible $s$ - wave scattering; for an outline of the derivation see Appendix \ref{sec:Unitary}.  It is empirically known that the in-plane resistivity of several layered materials follows the unitary scattering prediction to a precision of $\sim 30\%$. This is true of Sr$_{2}$RuO$_{4}$ \cite{kikugawa_non-fermi-liquid_2002}, La$_{2-x}$Sr$_{x}$CuO$_{4}$ and YBa$_{2}$Cu$_{3}$O$_{7-y}$ \cite{fukuzumi_universal_1996} in which the defects are introduced by substitution (Ti for Ru in Sr$_{2}$RuO$_{4}$, Zn for Cu in the cuprates), as well as of electron-irradiated YBa$_{2}$Cu$_{3}$O$_{7-y}$ \cite{rullier-albenque_universal_2000}. A comparison of our measurements with the unitary scattering limit is therefore simultaneously a comparison with other layered materials; a strong scattering suppression in PtCoO$_{2}$ would correspond to a strong deviation from the unitary scattering prediction.

\begin{figure}[t]
\includegraphics[scale=0.95]{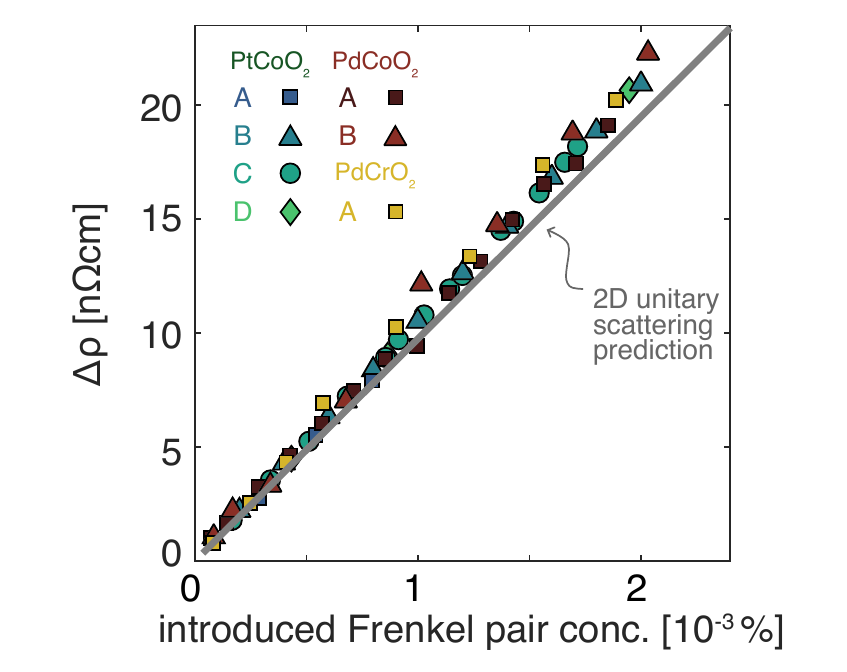}
\caption{\label{fig:Unitary} Resistivity increment of four PtCoO$_{2}$ samples,two PdCoO$_{2}$ samples and a PdCrO$_{2}$ sample  as a function of introduced Frenkel pair concentration compared to the unitary scattering prediction.}
\end{figure}

To test if this is the case, in Fig.\ \ref{fig:Unitary} we compare the resistivity increment in several samples of the three delafossites, PtCoO$_{2}$, PdCoO$_{2}$ and PdCrO$_{2}$, to the unitary scattering prediction. The introduced Pd Frenkel pair concentration in PdCoO$_{2}$ and PdCrO$_{2}$ is calculated assuming that the displacement energy for Pd is the same as that measured for Pt, $\unit[10]{eV}$, yielding a Pd Frenkel pair production cross-section of $\sigma_{Fp}\left(\text{Pd}, \unit[2.5]{MeV}\right)=\unit[315]{barn}$. The point defect concentration $n_{d}$ in equation \ref{eq:unitary} is taken to be equal to the introduced Frenkel pair concentration. The striking agreement of the resistivity increment and the unitary prediction demonstrates that the resistivity of delafossites is as sensitive to defects as that of other layered materials. This analysis conclusively shows that the enormous low-temperature mean free paths of the delafossite metals are caused by their unusual purity.

It is worth emphasising that the unitary scattering prediction does not depend on any free parameters, and the only fitting parameter leading to the conclusion drawn from Fig.\ \ref{fig:Unitary} is the displacement energy, which has been independently experimentally determined, as described in Sec.\ \ref{sec:Quant}. The analysis, however, equates the concentration of strong in-plane scatterers with that of Frenkel pairs, and assumes that the resistivity arises only from those in-plane scatterers. This cannot be exactly true, because each Frenkel pair consists of both an in-plane vacancy, and an interstitial, which is likely to be out of plane. The former are strong in-plane scatterers for which the unitary scattering is expected to be applicable, while the latter scatter more weakly. In practice, the resistivity is increased due to both of them, suggesting that the added resistivity per in-plane vacancy is smaller than suggested by the unitary scattering. However, even in the extreme and nonphysical case of out-of-plane interstitials adding as much resistivity as the in-plane vacancies, the resistivity would be suppressed by at most a factor of two from the unitary value. This in no way changes the conclusion that the extremely long mean free paths in delafossites arise mainly from their remarkable purity. 

The excellent agreement between the measurements on the three compounds indicates that the large difference in the resistivity increment as a function of electron dose shown in Fig.\ \ref{fig:ResisitivityIncrease} can be attributed solely to the lower differential cross-section for the formation of Pd defects, and that the resistivity of the three compounds is equally sensitive to defects. This allows for an estimate of the Frenkel pair concentration in the as grown crystals of PdCoO$_{2}$: the residual resistivity in the purest crystals reported to date is $8.1\unit{n\Omega cm}$ \cite{nandi_unconventional_2018}, corresponding to an estimated point defect concentration of $0.001\%$. 

\section{\label{sec:FirstPrinciples} First-principles calculations}

As discussed above, the irradiation studies indicate that the low resistivity of  PdCoO$_{2}$ and PtCoO$_{2}$ results from an unusually low defect density in the as-grown materials. In order to explore this theoretically, we use density functional theory (DFT) to calculate properties of Pd/Pt-related native defects in PdCoO$_2$ and PtCoO$_2$ including vacancies, interstitials, and Frenkel pairs. For computational details, see Appendix \ref{sec:ComputationalDetails}.

\subsection{\label{sec:Enthalpy}Formation energy comparison}

As a result of configurational entropy, there is always a finite concentration of native point defects in a crystal in thermodynamic equilibrium. Additional imperfections may exist as a result of kinetic barriers that prevent the system from reaching thermal equilibrium. We would like to determine whether the reduced defect concentrations observed in PdCoO$_2$ and PtCoO$_2$ are a result of the intrinsically higher formation energy of Pd/Pt-related defects, or from growth-specific kinetic factors.

The defect concentration is given by the formation free energy of the defect \cite{grabowski_ab_2009,freysoldt_first-principles_2014}. However, this quantity is computationally very intensive to determine accurately in complex materials. Thus our strategy will be to compare the formation energy (i.e., we neglect  electronic and vibrational entropy) of Pd-related defects in PdCoO$_2$ with those in Pd metal, in order to determine if they are significantly larger, and could explain the low defect concentrations in the delafossite metals. The formation energy of defect $X$ in a metal is given by

\begin{equation}
  \label{form}
  E_{\text{form}}[X]=E_{\text{tot}}[X]-\left(E_{\text{tot}}[{\text{bulk}}]+ \sum_in_i\mu_i\right).
\end{equation}

where $E_{\text{tot}}[X]$ is the energy of a supercell containing the
defect $X$, $E_{\text{tot}}[\text{bulk}]$ is the energy of a bulk cell
of corresponding size, $n_i$ is the number of atomic species $i$ added
or removed to create the defect, and $\mu_i$ is the chemical potential
of species $i$. $E_{\text{tot}}[X]$ and $E_{\text{tot}}[\text{bulk}]$
are calculated using DFT, while the chemical potential(s) $\mu_i$ are
specific to the experimental growth or annealing conditions. By
enforcing stability conditions, we can obtain theoretical limits on
$\mu$, which can provide guidance as to the possible experimental
conditions; for the calculations below, we assume Pd-rich (i.e., O-poor) conditions (see Appendix \ref{sec:ChemicalPotentials}).
 
In Table \ref{FormationEnergies} we  compare the energies of Pd vacancies, interstitials, and Frenkel pairs in PdCoO$_{2}$ with those in elemental Pd. Our formation energy for the Pd vacancy and self interstitial in Pd metal are consistent with previous calculations \cite{foiles_embedded-atom-method_1986,mattsson_calculating_2002, nazarov_vacancy_2012} and a slight underestimation compared to the experimental range of $1.5-1.85$ eV \cite{schaefer_investigation_1987,ullmaier_atomic_1991,  nazarov_vacancy_2012}. The general behaviour is similar between Pd metal and PdCoO$_2$, with vacancies expected to be the dominant defect. The formation energy of interstitials in PdCoO$_{2}$ is significantly larger that in Pd; however, in both cases, it is not expected that interstitials will be present in large concentrations. The large formation energy of interstitials is expected in close-packed metals. For  PdCoO$_{2}$ it is likely due to the short Pd-O bond lengths of $\sim$ 2 $\unit{\AA}$ (\emph{cf.} 2.8 $\unit{\AA}$ for the Pd-Pd bonds in PdCoO$_2$),  forcing the interstitial atom unfavourably
close to the Pd layer (for a more detailed discussion of the interstitial structure see Appendix \ref{sec:InterstitialStructure}).

The Frenkel pair formation energy in both cases is approximately the sum of the vacancy and interstitial energies. Thus, there is not a significant energy gain from having the interstitial in the vicinity of the vacancy, and we find that in the delafossites, the interstitial needs to be placed several lattice sites away to prevent it from relaxing back to the vacancy site (see Sec.~\ref{sec:DisplacementCalc}).

We find similar behavior between PtCoO$_2$ (not shown) and PdCoO$_2$. We therefore find no evidence that the exceptionally low defect concentrations seen in  as-grown delafossites are a result uncharacteristically high formation energies for Pd/Pt defects in those materials.

\begin{table}[h]
  \caption{\label{FormationEnergies} Formation energies in eV for Pd-related native defects in PdCoO$_2$ and Pd metal. In parentheses are previous DFT-GGA calculations}
\begin{ruledtabular}
  \label{form_eng}
  \begin{tabular}{ccc}
    defect & PdCoO$_2$  & Pd  \\
    \hline
    Pd vacancy & 0.88    & 1.26 (1.20\footnotemark[1],1.19\footnotemark[2]) \\
    Pd interstitial & 6.28 & 3.82 (3.43\footnotemark[3]) \\
    Pd Frenkel pair & 7.11 &  5.11 \\
\end{tabular}
\end{ruledtabular}
\footnotetext[1]{Ref.~\onlinecite{mattsson_calculating_2002}}
\footnotetext[2]{Ref.~\onlinecite{nazarov_vacancy_2012}}
\footnotetext[3]{Ref.~\onlinecite{foiles_embedded-atom-method_1986}}
\end{table}

\subsection{\label{sec:DisplacementCalc}Displacement energy}

Of course, the formation energy of Frenkel pairs calculated in Sec.~\ref{sec:Enthalpy} does not apply to the nonequlibrium creation of defects via irradiation; the important quantity in this case is the displacement energy. To obtain a theoretical estimate of the displacement energy, we calculate the adiabatic energy barrier for a Pt/Pd atom to be displaced from its  position in the lattice towards an interstitial site using the nudged elastic band (NEB) method. 

\begin{figure}[h]
  \includegraphics[scale=1]{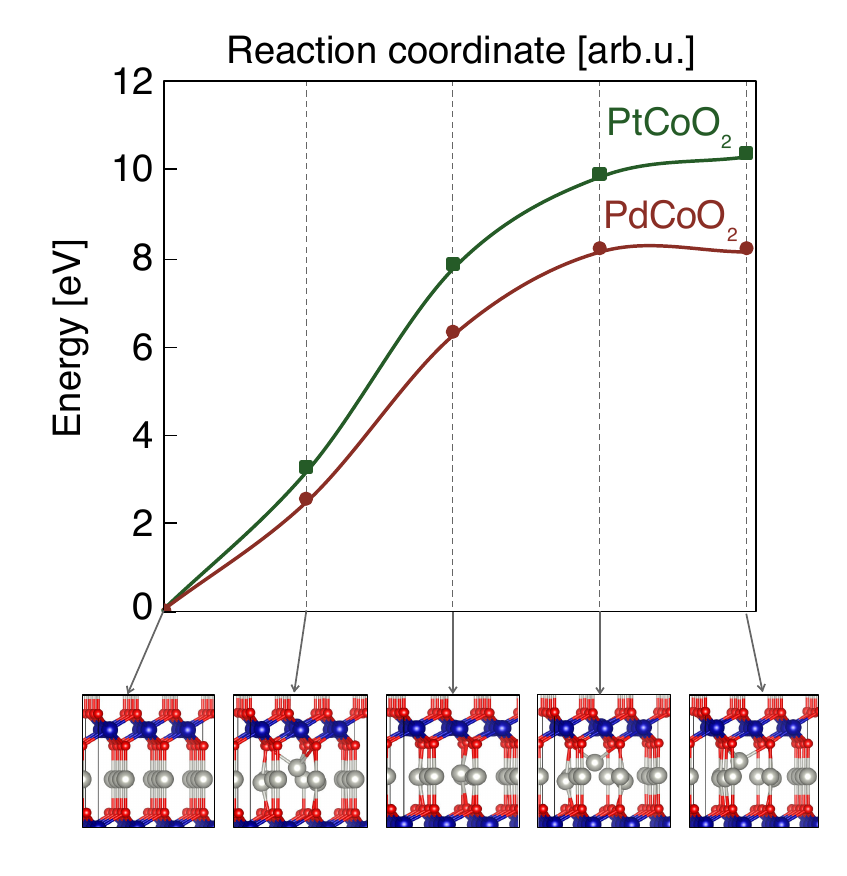}
  \caption{\label{FrenkelNEB} Adiabatic barrier for Frenkel pair
    formation in PdCoO$_2$ and PtCoO$_2$, from nudged elastic band DFT
    calculations. }
\end{figure}

The results are shown in Fig.~\ref{FrenkelNEB}, where the initial structure is the ideal bulk structure, and the final structure is that of the relaxed Frenkel pair. We can see that for both PdCoO$_2$ and PtCoO$_2$, there is essentially no additional kinetic barrier to the formation: the energy required to adiabatically create the Frenkel pair is approximately equal to its formation enthalpy, i.e \unit[$\sim8$]{eV} (see Appendix \ref{sec:ComputationalDetails}) in PdCoO$_{2}$ and \unit[$\sim10$]{eV} in PtCoO$_{2}$. It is important to note that the total energy of the barrier should be considered as a lower bound for the displacement energy, as it allows for all degrees of freedom perpendicular to the reaction coordinate to relax, which will likely not occur in the actual Mott scattering process. The results are therefore fully consistent with the value deduced from experiment of \unit[10]{eV}. 

\section{\label{sec:Transport} Influence of defects on transport properties}

All of the experiments and analysis shown so far address the question of the influence of defects on the low-temperature resistivity of delafossite metals. It is often implicitly assumed that the various contributions to resistivity are additive, obeying the so-called Matthiessen's rule. Measuring the transport properties of a sample before and after electron irradiation offers a unique opportunity to test this assumption in a controlled manner. Such a study is, however, by no means guaranteed to succeed. First of all, as the sample is warmed up from the irradiation temperature of $\unit[22]{K}$ the defects become more mobile and may recombine, therefore removing any added resistivity. Even if the defects remain in the sample by room temperature, they may slowly anneal, leading to time-dependent resistivity. In PtCoO$_{2}$ the added resistivity does decrease during the initial warm up to room temperature to $\sim 65\%$ of its original value. However, the remaining resistivity stays unchanged at room temperature and during subsequent thermal cycles to low temperatures (for more details see Appendix \ref{sec:Annealing}), enabling a controlled and reproducible study of the disorder influence on transport. Such a study is beyond the scope of this work, but we show below two intriguing results which point to its broad significance. 

\begin{figure}[h]
\includegraphics[scale=1.1]{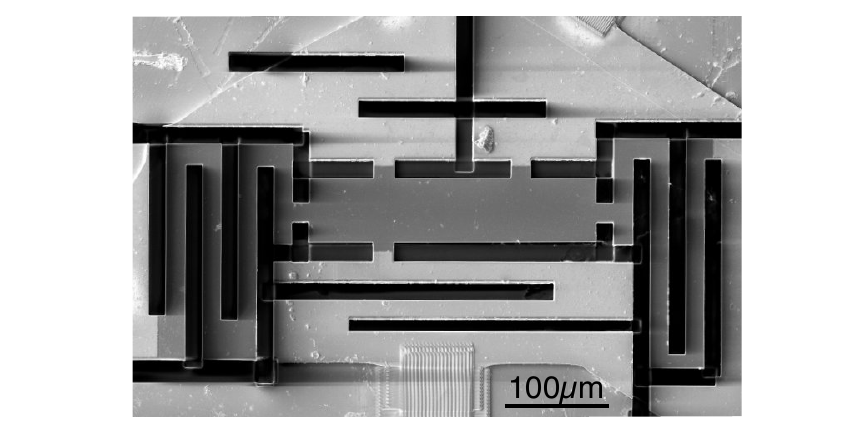}
\caption{\label{fig:HallStructure} An SEM image  of  a
microstructured PtCoO$_{2}$ sample used for the ex-situ measurements of the temperature and magnetic field dependent resistivity.}
\end{figure}

In Fig.\ \ref{fig:HallStructure} we show an SEM image of a sample we have structured for the  temperature and magnetic field dependent measurements. Its well-defined bar geometry enables a precise determination of its resistivity, which prior to irradiation agrees quantitatively with the published data \cite{nandi_unconventional_2018}. The estimated defect concentration of the sample before irradiation was $0.0046\%$, while after the irradiation to a total dose of $\unit[188]{mC/cm^2}$ by electrons of $\unit[2.5]{MeV}$ kinetic energy, and a subsequent warm-up to room temperature, it was increased by nearly an order of magnitude, to $0.038\%$. 

\begin{figure}[t]
  \includegraphics[scale=1]{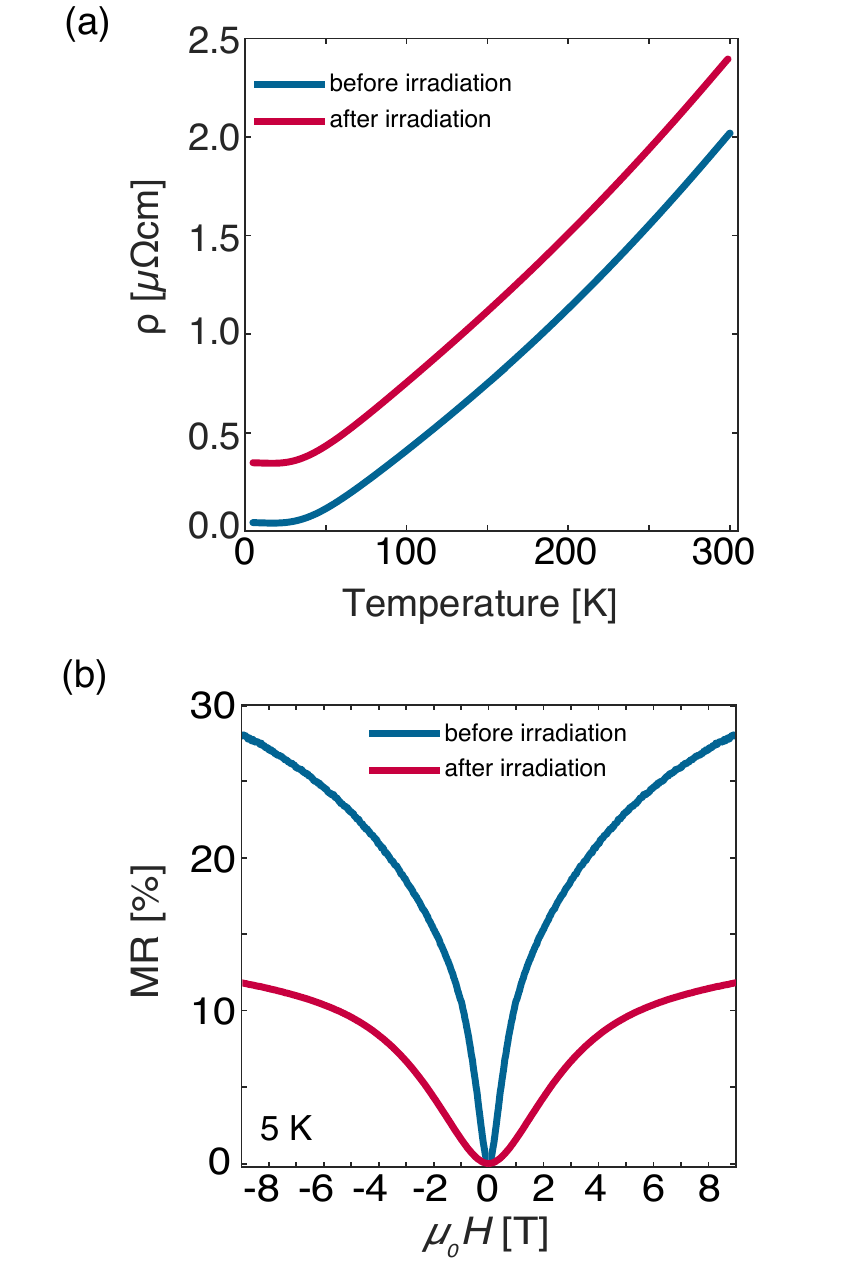}
  \caption{\label{Transport} The (a) temperature-dependent resistivity and (b) magnetoresistance ($MR=\left(\rho\left(H\right)-\rho\left(0\right)\right)/\rho\left(0\right)$) of the sample shown in Fig.\ \ref{fig:HallStructure} before and after irradiation.}
\end{figure}

In Fig.\ \ref{Transport} we show the resistance of the sample shown in  Fig.\ \ref{fig:HallStructure} as a function of temperature (Fig.\ \ref{Transport}a) and magnetic field  (Fig.\ \ref{Transport}b), before and after irradiation. The temperature dependence of resistivity remains nearly unchanged, with the defects adding an approximately temperature-independent offset, indicating a good agreement with Matthiessen's rule. 

Introduced defects reduce the overall scale of magnetoresistance by a factor of $2.3$. Again, this reduction can be qualitatively understood  within the semi-classical picture of magnetoresistance, assuming independent scattering events. Finite magnetoresistance indicates the presence of more than one length scale governing transport, and is proportional to the square of their difference \cite{ziman_principles_1972}. In a single band material of nearly constant Fermi velocity \cite{nandi_unconventional_2018} this suggests at least two different scattering rates at different points of the Fermi surface, $\tau_{1}$ and $\tau_{2}=\alpha \tau_{1}$; the magnetoresistance is proportional to $\left(\tau_{1}-\tau_{2}\right)^2=(1-\alpha)^2\tau_{1}^2$.  If Matthiessen's rule is obeyed and the defect scattering exhibits the same $\textbf{k}$ - dependence as $\tau_{1}$ and $\tau_{2}$ (certainly a good approximation at low temperatures), the scattering times are changed by defects into $\tau_{1d}$ and $\tau_{2d}=\alpha \tau_{1d}$. The magnetoresistance is then proportional to $\left(\tau_{1d}-\tau_{2d}\right)^2=(1-\alpha)^2\tau_{1d}^2$. The defects reduce the scattering  times  ($\tau_{1d}<\tau_{1}$), so this represents a reduction in magnetoresistance compared to the case without the added defects. Therefore, regardless of the cause of the different scattering rates across the Fermi surface, adding a source of scattering to all the states decreases magnetoresistance, as observed. 

Although at first sight these observations are not unusual, a more detailed analysis offers additional insight, as discussed in the following section. 

\section{\label{sec:Discussion} Discussion}

\begin{figure}[b]
\includegraphics[scale=1]{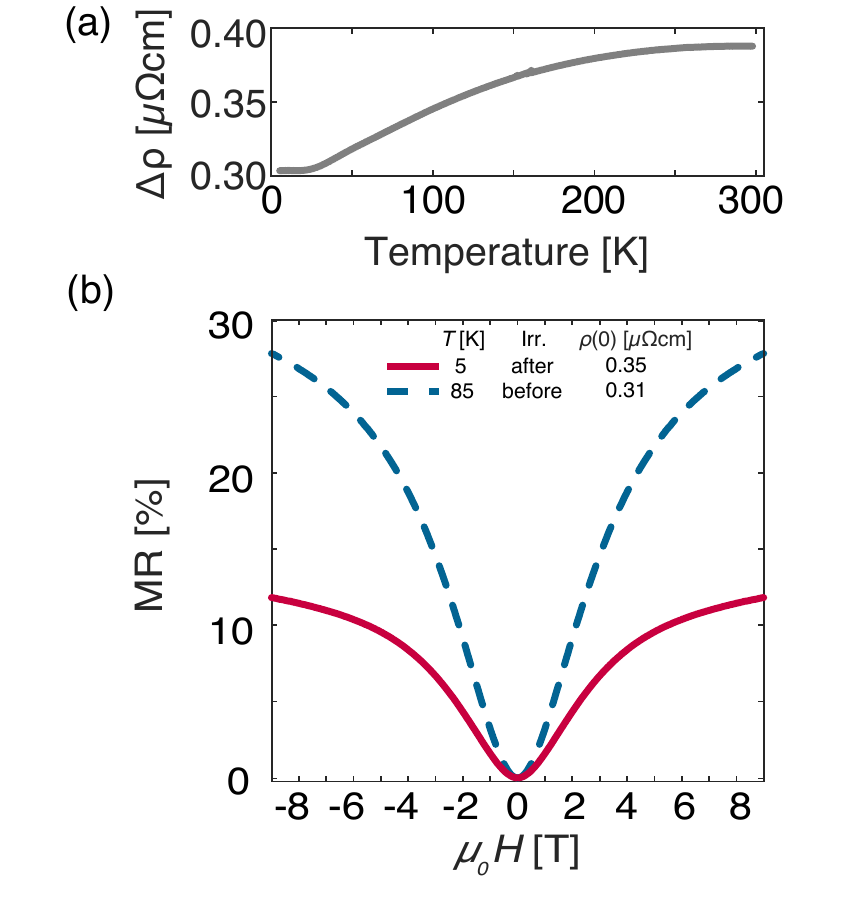}
\caption{\label{TransportDetail} (a) The difference of the temperature dependent resistivity after and before the irradiation. (b) The magnetoresistance ($MR=\left(\rho\left(H\right)-\rho\left(0\right)\right)/\rho\left(0\right)$)  measured at $\unit[5]{K}$  after irradiation compared to that measured at $\unit[85]{K}$ before irradiation (dashed line). The zero field resistivity (legend) was the same within  $10\%$ in the two measurements.}
\end{figure}

The experimental data and analysis presented in this paper provide very strong evidence in favour of a remarkable experimental fact: the extremely long low-temperature mean free paths of metallic delafossites are mainly due to an exceptional level of crystalline perfection of the Pd or Pt planes in which the conduction takes place.  We are able to deduce point defect levels as low as $0.001\%$ in the best as-grown crystals.  

The level of control of defect densities that we have demonstrated, combined with the data quality obtainable from the microstructures that we have used, will also open paths to new discoveries.  For example, we will be able to carry out a stringent examination of subtle effects that challenge understanding of scattering of electrons in solids. Although the agreement with Matthiessen’s rule, shown in Fig.\ \ref{Transport}a, looks excellent on first inspection, more detailed analysis demonstrates a systematic, temperature-dependent deviation, as seen in the plot of the difference between the temperature dependent resistivity of the irradiated and pristine sample (Fig.\ \ref{TransportDetail}a).  This suggests that the scattering from the in-plane defects that we have introduced is not entirely independent of electron-phonon scattering. The data shown in Fig.\ \ref{TransportDetail}b offer further information on the interplay of the two sources of scattering.  At \unit[5]{K} after irradiation, the resistivity is almost identical to that at \unit[85]{K} in the as-grown crystal, so the resistive mean free paths averaged around the Fermi surface are also very similar.  However, the magnetoresistance, which is sensitive to the $\textbf{k}$-dependence of the mean free path around the Fermi surface \cite{harris_violation_1995}, differs in scale by over a factor of two.  The Fermi velocity is $\textbf{k}$-independent to within a few per cent \cite{nandi_unconventional_2018}, so mean free path variations must be due to variations in the scattering rates, indicating that the $\textbf{k}$-dependence of the scattering rate depends on the scattering mechanism. These observations strongly motivate future work to study the magnetotransport as a function of temperature and defect density, and the extent to which such data can be modelled using modern quantum transport theories \cite{groth_kwant:_2014} and calculations of the $\textbf{k}$-dependence of the electron-phonon interaction \cite{garcia_optoelectronic_2019}.

Another avenue for future work that the current findings will stimulate is closer investigation of other materials that crystallize in the delafossite structure.  Is the crystalline perfection that we have uncovered unique to the Pt and Pd layers of PdCoO$_{2}$, PdCrO$_{2}$ and PtCoO$_{2}$, or might it occur in other delafossites as well?  In particular, it will be interesting to investigate the crystalline purity of non-metals from this structural class; this might be a route to the development of a new family of high-mobility semiconductors.  We note, however, that the in-plane purity is sufficient to achieve a high conductivity in delafossite \emph{metals}, because the out-of-plane impurities are very efficiently screened; this is proven by the data shown in Fig.\ \ref{fig:ResisitivityIncrease} and their analysis in Fig.\ \ref{fig:Unitary}. In contrast, a high-mobility semiconducting delafossite would require crystalline perfection in all its layers, as the screening of out-of-plane defects would be less efficient \cite{coleridge_small-angle_1991}. While at present the level of disorder in the Co layers of PdCoO$_{2}$ and PtCoO$_{2}$ is not known, we confirm it to be below the resolution of  electron microscopy (Fig.\ \ref{fig:TEM}). Combined with the remarkable purity of the Pd and Pt layers we have uncovered, this motivates careful research on close structural relatives of these fascinating compounds.

\section{\label{sec:Conclusions} Conclusions}

In this paper we have used electron beam irradiation to introduce Frenkel pairs into the crystal structure of the delafossite metals PdCoO$_{2}$, PdCrO$_{2}$ and PtCoO$_{2}$, to investigate the reason for their extremely long low-temperature mean free paths.  By studying three compounds in which the cations change between Pd and Pt, and Co and Cr, we have demonstrated empirically that the resistivity is sensitive to defects in the Pd/Pt layers.  Varying the energy of the incident electrons between $0.8$ and $\unit[2.5]{MeV}$ produces data that can be fitted very well with Mott scattering theory with only one free fit parameter, which is independently confirmed by first-principles electronic structure calculations. The excellent fit to the Mott theory enabled quantitative measurement of the Frenkel pair production cross-sections and hence the determination of the defect density of each irradiated crystal. The additional scattering thus introduced is in quantitative agreement with so-called unitary scattering, and the data allowed a reliable estimate of the density of point defects in as-grown crystals.  This is as low as $0.001\%$, proving that the main reason for the long mean free paths is a level of crystalline perfection rarely seen in multi-component oxides.  These findings raise important questions about the levels of purity potentially attainable in non-metallic delafossites, and motivate further investigation of this intriguing class of materials.

\begin{acknowledgments} We thank C. Hooley, P.D.C. King and R. Moessner for useful discussions, and A. Georges for suggesting complementing the experiments with the first principles calculations. We acknowledge support from the Max Planck Society. V.S. and P.H.M. acknowledge EPSRC for PhD studentship support through grant number EP/L015110/1. E.Z. acknowledges support from the IMPRS for the Chemistry and Physics of Quantum Materials. 
M. Konczykowski acknowledges support from SIRIUS irradiation facility with project EMIR 2019 18-7099. Electron Microscopy at Cornell is supported by the US National Science Foundation (Platform for the Accelerated Realization, Analysis, and Discovery of Interface Materials (PARADIM)) under Cooperative Agreement No. DMR-1539918, and  DMR-1719875.

\end{acknowledgments}

\appendix

\section{\label{sec:Unitary}{Unitary scattering limit in 3D and 2D}}

Within the Drude model the resistivity of a 3D material is given by: 

\begin{equation}
\label{eq:3D resistivity}
\rho^{3D}=\frac{m_{eff}}{ne^2}\frac{1}{\tau_{tr}}=\frac{m_{eff}}{ne^2}n_{d}v_{F}\left<\sigma_{imp}\right>,
\end{equation}
where $m_{eff}$ is the effective mass of the carriers, $n$ their concentration, $v_{F}$ the Fermi velocity, $n_{d}$ the defect concentration and $\tau_{tr}$ the transport lifetime.  $\left<\sigma_{imp}\right>$ is the cross-section for scattering of electrons off impurities, weighted by the $\left(1-\cos\vartheta\right)$ term accounting for the larger contribution to resistivity of backscattering compared to small angle scattering. 

If $\left<\sigma_{imp}\right>$  is calculated using partial wave analysis assuming a central potential, it is found to be equal to \cite{solyom_fundamentals_2008}:
\begin{equation}
\label{eq:imp_Cross_Section}
\left<\sigma_{imp}\right>=\frac{4\pi}{k^2}\sum\limits_{l=0}^\infty\left(l+1\right)\sin^2\left(\delta_l-\delta_{l+1}\right),
\end{equation}
 where $k$ is the wavevector of the electrons, and $\delta_{l}$ is the phase shift associated with the $l$-th Legendre polynomial comprising the electron wave function. The unitary limit corresponds to the strongest possible $s$-wave scattering, in which $\delta_{0}=\pi/2$ and all other $\delta_{l}=0$. 
 The resistivity in the unitary limit is then given by: 
 
 \begin{equation}
\label{eq:3D_Unit}
\rho_{unit}^{3D}=\frac{4\pi\hbar n_{d}}{n e^2k_{F}},
\end{equation}
where $k_{F}$ is the Fermi wavevector. 

Resistivity of a two-dimensional system is given by an expression analogous to equation \ref{eq:3D resistivity}, but the 3D scattering area $\left<\sigma_{imp}\right>$ has to be replaced by a 2D scattering length $\left<\lambda_{imp}\right>$, given by \cite{lapidus_quantummechanical_1982}:

\begin{equation}
\label{eq:imp_Cross_Section2}
\left<\lambda_{imp}\right>=\frac{4}{k}\left(1+2\sum\limits_{m=1}^\infty\sin^2\delta_{m}\right).
\end{equation}
In the unitary limit the scattering length is reduced to $\left<\lambda_{imp}\right>=4/k$; it is therefore comparable to the Fermi wavelength. The resistivity is given by: 
\begin{equation}\label{eq:unitary2D}
\rho_{unit}^{2D}=\frac{4\hbar}{e^{2}}\frac{n_{d}}{n}, 
\end{equation}
where we used $m_{eff}v_{F}=\hbar k_{F}$. 

\section{\label{sec:ComputationalDetails}{Computational Details}}

\begin{figure}[b]
  \includegraphics[width=\columnwidth]{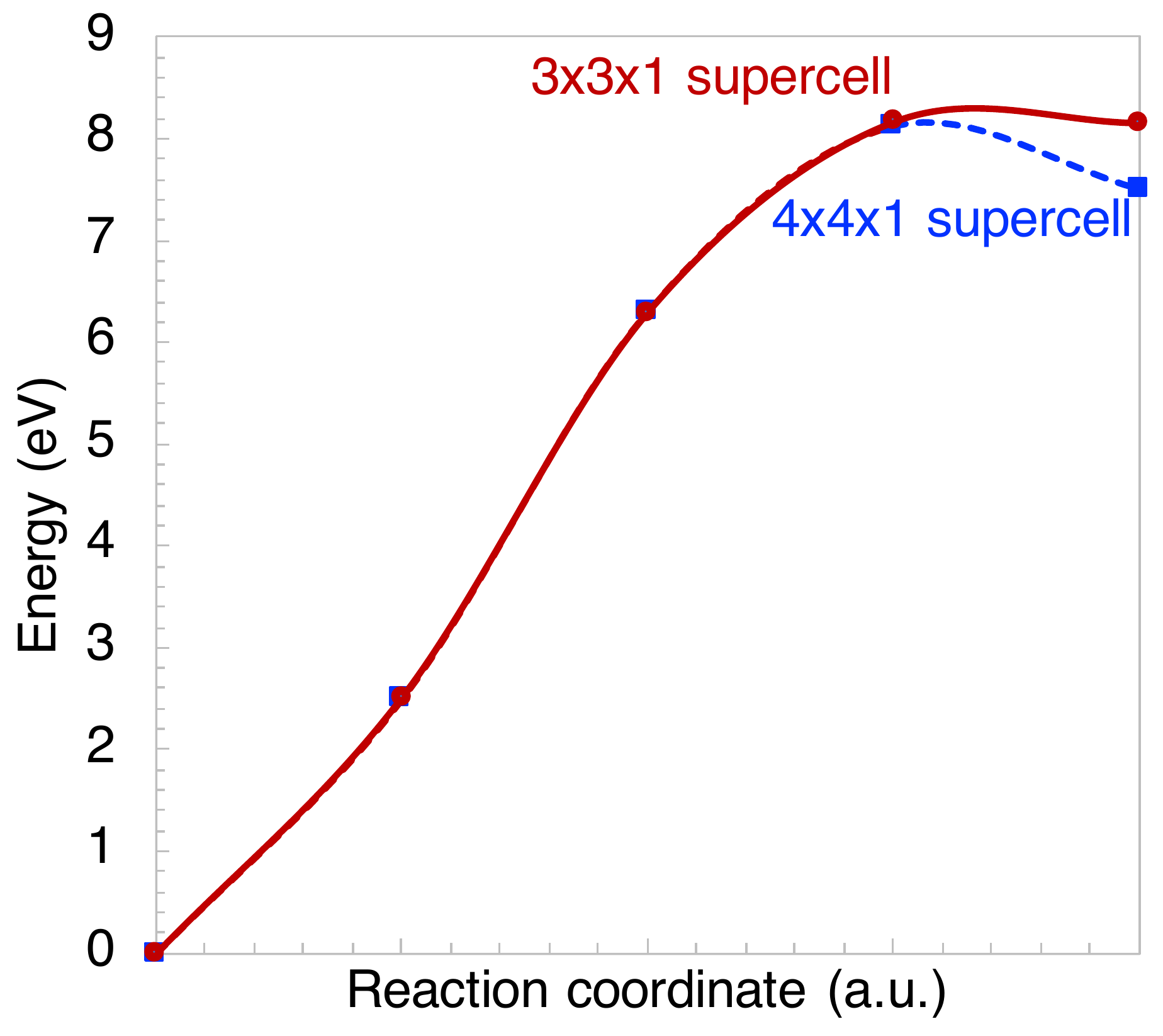}
  \caption{\label{Frenkel_sc} Frenkel barrier dependence on supercell size for PdCoO$_2$.  }
\end{figure}

Density functional theory (DFT) calculations were performed using the
PBE generalized gradient functional \cite{perdew_generalized_1996}, implemented in
the {\sc vasp} code \cite{kresse_efficient_1996}. It was shown in
Ref.~\cite{ong_origin_2010} that the experimental electronic structure
\cite{kushwaha_nearly_2015} can be accurately reproduced with GGA+$U$ with the
Hubbard $U = 4$ eV on the Pd/Pt $d$-states, as well as spin-orbit
coupling (SOC) included; however the addition of a Hubbard $U$ and SOC have relatively minor effects on the electronic structure, and were thus neglected in this study. 
The experimental lattice constants
(for the conventional cell) of $a=2.830$ {\AA}, $c=17.743$ {\AA}
(PdCoO$_2$) \cite{shannon_chemistry_1971} and $a=2.823$ {\AA}, $c=17.808$ {\AA}
(PtCoO$_2$) \cite{kushwaha_nearly_2015} were used throughout.

A 520 eV
cutoff for the plane-wave basis set was used.
A supercell of $5\times 5\times 1$ conventional cells (300 atoms) was
used to calculate the formation energies of defects in the delafossites, with a $2\times2\times2$ gamma-centered $k$-mesh. For Pd metal, a $4\times4\times4$ supercell and $4\times4\times4$ $k$-mesh was used.
Diffusion barriers and Frenkel pair formation barriers were
calculated using the nudged elastic band (NEB) method; due to the computational intensive nature of these calculations, a $3\times3\times1$ cell (and $3\times3\times1$ $k$-mesh) was used. The relatively small in-plane area of the supercell results in a overestimation of the energy of the separated Frenkel pair, as we demonstrate in Fig.~\ref{Frenkel_sc} for PdCoO$_2$, but does not affect our estimate of the barrier for Frenkel pair formation.

\section{\label{sec:ChemicalPotentials}{Limits on the chemical potentials}}

Under thermodynamic equilibrium the chemical potentials in
Eq.~(\ref{form}) are limited by the stability of the bulk material. We
illustrate this for the case of PdCoO$_2$. If we reference the
chemical potentials to the corresponding elemental phases (i.e., the
elemental metals for Pd and Co, and an isolated O$_2$ molecule for O),
the stability condition of PdCoO$_2$ can be written
\cite{freysoldt_first-principles_2014}
\begin{equation}
  \label{chempot}
  \mu_{\text{Pd}}+\mu_{\text{Co}}+2\mu_{\text{O}}=\Delta H_f(\text{PdCoO}_2)
\end{equation}
where $\Delta H_f(\text{PdCoO}_2)$ is the enthalpy of formation of
PdCoO$_2$. Thus, only two chemical potentials can be varied
independently; fixing those values will specify the third.

\begin{figure}[h]
  \includegraphics[width=\columnwidth]{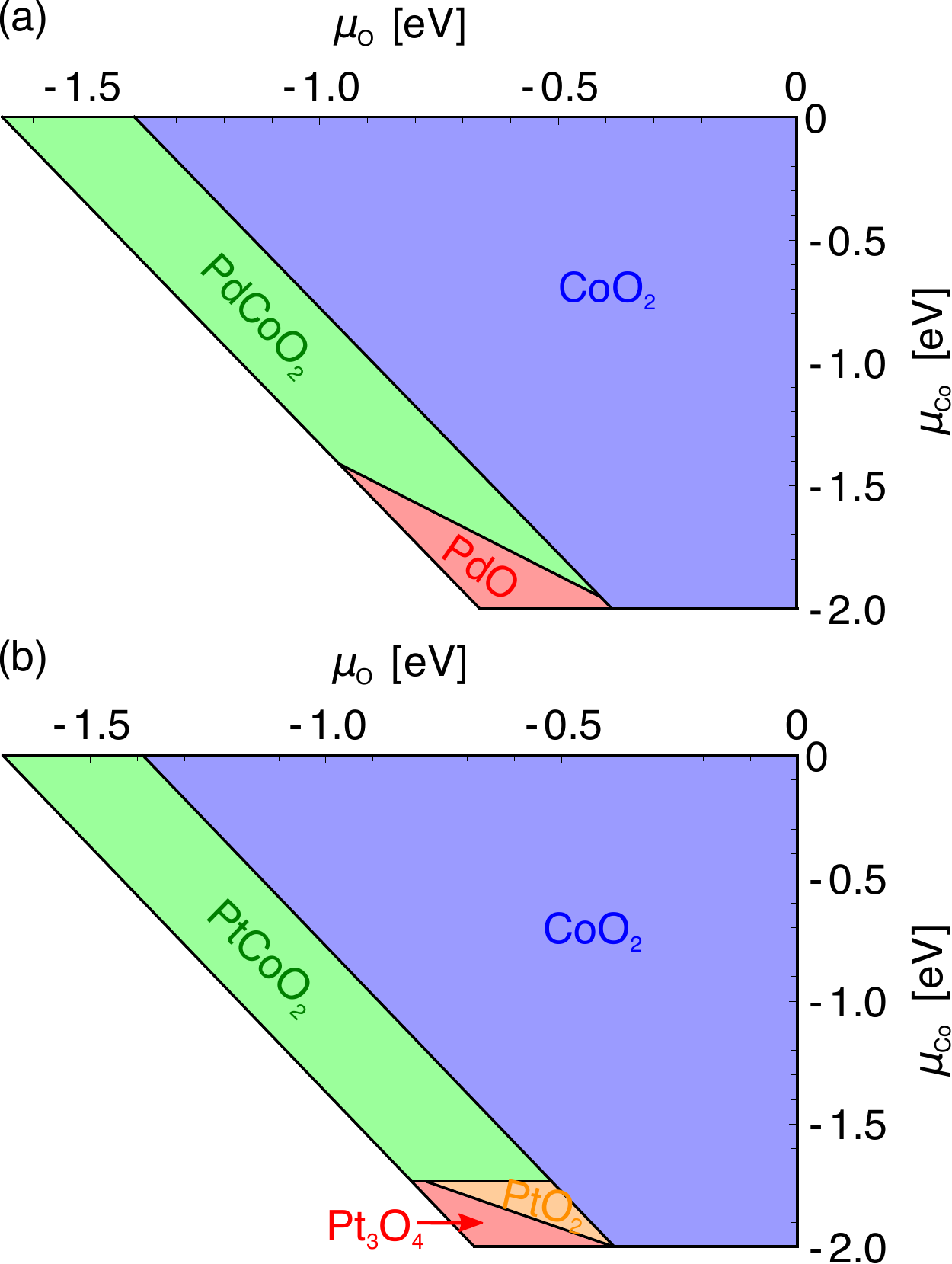}
  \caption{\label{mu_limit} $T=0$ stability diagram for (a) PdCoO$_2$
    and (b) PtCoO$_2$ with respect to O and Co chemical
    potentials. Green region corresponds to chemical potentials where
    the delafossites are stable.  }
\end{figure}

The range of chemical potentials during growth will be further limited
by the formation of other phases with different stoichiomentry. In
particular, we find that if conditions are too Co-poor and/or O-poor,
PdO will form over PdCoO$_2$ (equivalently, Pt$_3$O$_4$ over
PtCoO$_2$), and if the conditions are too O-rich and/or Co-rich,
CoO$_2$ will form. There is also a small O-rich, Co-poor region where PtO$_2$ is the most stable. These limits are shown in Fig.~\ref{mu_limit}. In general, we see that CoO$_2$ limits stability of the delefossites to fairly Pd/Pt-rich conditions, which will generally increase the formation energy of Pd/Pt vacancies, which are the dominant defects in this system.
For calculating formation energies in Sec.~\ref{sec:Enthalpy}, we assume metal-rich conditions, corresponding to $\mu_{\text{Pd}}=\mu_{\text{Co}}=0$, giving $\mu_{\text{O}}=-1.67$ eV for PdCoO$_2$.

\section{\label{sec:InterstitialStructure}{Interstitial Defect Structure}}

The most stable interstitial structure that we find in PdCoO$_{2}$ is for the interstitial atom (Pd$_i$) to bond to two O in the layer above, resulting in significant in-plane displacements of two Pd atoms. The bond lengths between the interstitial and the two oxygen atoms are not equal (1.96 and 2.19 \AA), while one of  the nearby Pd is significantly more displaced from its original site than the other. The bond lengths between the Pd$_i$ and the nearest two Pd atoms are approximately equal (2.34 \AA). We can understand this
structure as a balance between the Pd$_i$ attempting to form
equilibrium length bonds with the two O atoms above, and the two Pd
atoms below. 

\section{\label{sec:Annealing}{Stability of defects}}

\begin{figure}[b]
\includegraphics[width=\columnwidth]{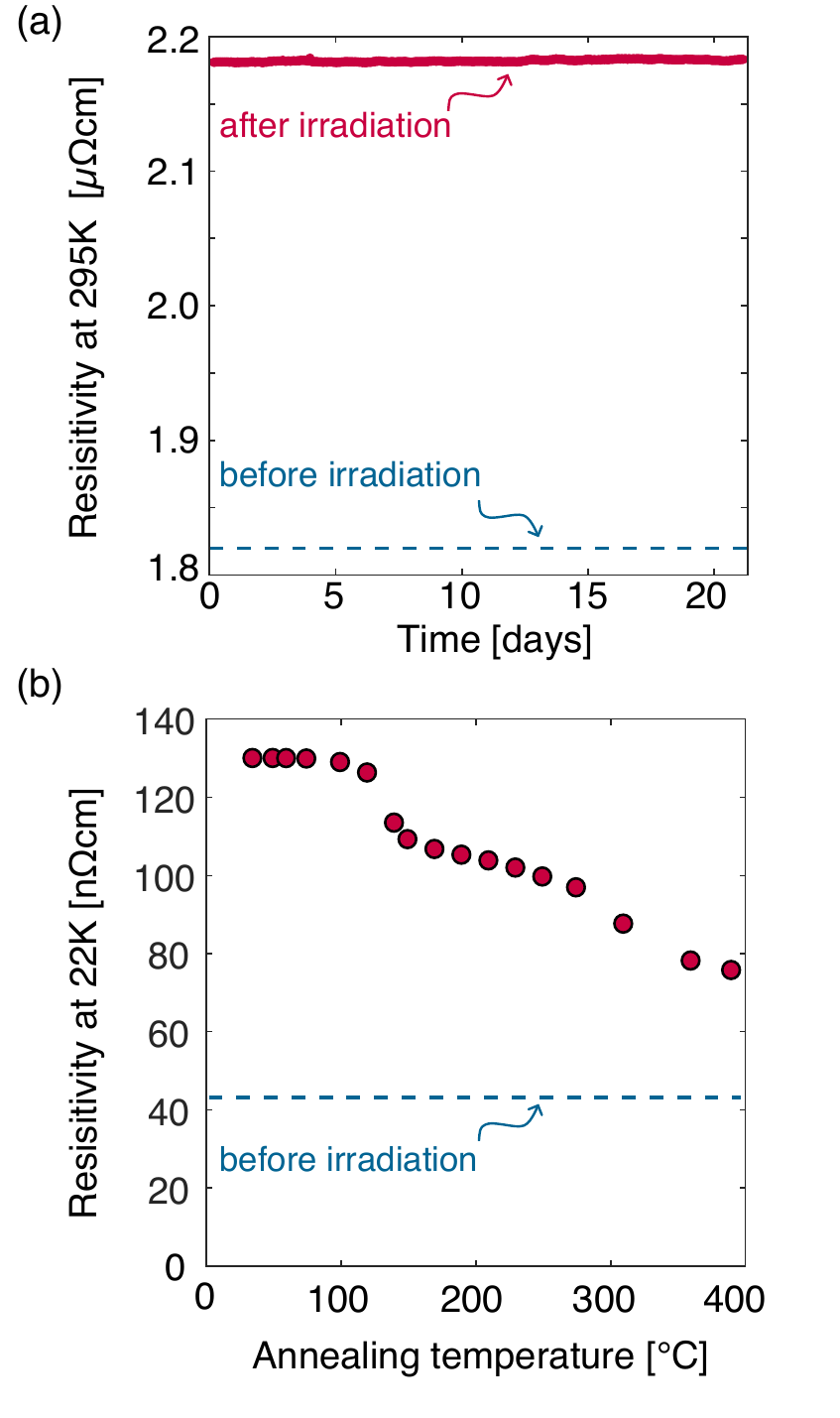}
\caption{\label{annealing} (a) The room temperature resistivity of an irradiated PtCoO$_{2}$ sample measured over the course of twenty days. (b) The resistivity of an irradiated PtCoO$_{2}$ sample measured at $\unit[22]{K}$ as a function of the annealing temperature, at which the sample was kept for $\unit[30]{min}$.}
\end{figure}

To check whether the defects added by irradiation are stable after the initial warm-up to room temperature, we measured the room-temperature resistivity of an irradiated PtCoO$_{2}$ sample continuously over the course of twenty days. As shown in Fig.\ \ref{annealing}a, no change of resistivity was observed during this time, indicating that the defects which remain in the sample after the initial warm-up to room temperature are indeed stable. Consistent with this, we observed no difference in the measurements of the temperature-dependent resistivity taken several months apart. 

At higher temperatures, however, the defects become mobile. To investigate this we have warmed up an irradiated PtCoO$_{2}$ sample to a series of annealing temperatures between $\unit[35]{\degree C}$ and  $\unit[390]{\degree C}$, and kept the sample at each of them for $\unit[30]{min}$. After each annealing step, the temperature dependent resistivity was measured. In Fig.\ \ref{annealing}b we show the resisitivity at $\unit[22]{K}$ as a function of annealing temperature. Two clear steps can be seen, indicating two characteristic energies for migration of different types of defects. The fact that the added resisitivity is decreased by $35\%$ after warming up to room temperature indicates that one or more such steps also occur between $\unit[22]{K}$ and room temperature.

\bibliography{MyLibrary}

\begin{thebibliography}{56}%
\makeatletter
\providecommand \@ifxundefined [1]{%
 \@ifx{#1\undefined}
}%
\providecommand \@ifnum [1]{%
 \ifnum #1\expandafter \@firstoftwo
 \else \expandafter \@secondoftwo
 \fi
}%
\providecommand \@ifx [1]{%
 \ifx #1\expandafter \@firstoftwo
 \else \expandafter \@secondoftwo
 \fi
}%
\providecommand \natexlab [1]{#1}%
\providecommand \enquote  [1]{``#1''}%
\providecommand \bibnamefont  [1]{#1}%
\providecommand \bibfnamefont [1]{#1}%
\providecommand \citenamefont [1]{#1}%
\providecommand \href@noop [0]{\@secondoftwo}%
\providecommand \href [0]{\begingroup \@sanitize@url \@href}%
\providecommand \@href[1]{\@@startlink{#1}\@@href}%
\providecommand \@@href[1]{\endgroup#1\@@endlink}%
\providecommand \@sanitize@url [0]{\catcode `\\12\catcode `\$12\catcode
  `\&12\catcode `\#12\catcode `\^12\catcode `\_12\catcode `\%12\relax}%
\providecommand \@@startlink[1]{}%
\providecommand \@@endlink[0]{}%
\providecommand \url  [0]{\begingroup\@sanitize@url \@url }%
\providecommand \@url [1]{\endgroup\@href {#1}{\urlprefix }}%
\providecommand \urlprefix  [0]{URL }%
\providecommand \Eprint [0]{\href }%
\providecommand \doibase [0]{https://doi.org/}%
\providecommand \selectlanguage [0]{\@gobble}%
\providecommand \bibinfo  [0]{\@secondoftwo}%
\providecommand \bibfield  [0]{\@secondoftwo}%
\providecommand \translation [1]{[#1]}%
\providecommand \BibitemOpen [0]{}%
\providecommand \bibitemStop [0]{}%
\providecommand \bibitemNoStop [0]{.\EOS\space}%
\providecommand \EOS [0]{\spacefactor3000\relax}%
\providecommand \BibitemShut  [1]{\csname bibitem#1\endcsname}%
\let\auto@bib@innerbib\@empty
\bibitem [{\citenamefont {Gardner}\ \emph {et~al.}(2016)\citenamefont
  {Gardner}, \citenamefont {Fallahi}, \citenamefont {Watson},\ and\
  \citenamefont {Manfra}}]{gardner_modified_2016}%
  \BibitemOpen
  \bibfield  {author} {\bibinfo {author} {\bibfnamefont {G.~C.}\ \bibnamefont
  {Gardner}}, \bibinfo {author} {\bibfnamefont {S.}~\bibnamefont {Fallahi}},
  \bibinfo {author} {\bibfnamefont {J.~D.}\ \bibnamefont {Watson}},\ and\
  \bibinfo {author} {\bibfnamefont {M.~J.}\ \bibnamefont {Manfra}},\ }\bibfield
   {title} {{\selectlanguage {english}\bibinfo {title} {Modified {MBE} hardware
  and techniques and role of gallium purity for attainment of two dimensional
  electron gas mobility {\textgreater}35 x
  10$^{\textrm{6}}$cm$^{\textrm{2}}$/{Vs} in {AlGaAs}/{GaAs} quantum wells
  grown by {MBE}}},\ }\href {https://doi.org/10.1016/j.jcrysgro.2016.02.010}
  {\bibfield  {journal} {\bibinfo  {journal} {Journal of Crystal Growth}\
  }\textbf {\bibinfo {volume} {441}},\ \bibinfo {pages} {71} (\bibinfo {year}
  {2016})}\BibitemShut {NoStop}%
\bibitem [{\citenamefont {Banszerus}\ \emph {et~al.}(2016)\citenamefont
  {Banszerus}, \citenamefont {Schmitz}, \citenamefont {Engels}, \citenamefont
  {Goldsche}, \citenamefont {Watanabe}, \citenamefont {Taniguchi},
  \citenamefont {Beschoten},\ and\ \citenamefont
  {Stampfer}}]{banszerus_ballistic_2016}%
  \BibitemOpen
  \bibfield  {author} {\bibinfo {author} {\bibfnamefont {L.}~\bibnamefont
  {Banszerus}}, \bibinfo {author} {\bibfnamefont {M.}~\bibnamefont {Schmitz}},
  \bibinfo {author} {\bibfnamefont {S.}~\bibnamefont {Engels}}, \bibinfo
  {author} {\bibfnamefont {M.}~\bibnamefont {Goldsche}}, \bibinfo {author}
  {\bibfnamefont {K.}~\bibnamefont {Watanabe}}, \bibinfo {author}
  {\bibfnamefont {T.}~\bibnamefont {Taniguchi}}, \bibinfo {author}
  {\bibfnamefont {B.}~\bibnamefont {Beschoten}},\ and\ \bibinfo {author}
  {\bibfnamefont {C.}~\bibnamefont {Stampfer}},\ }\bibfield  {title} {\bibinfo
  {title} {Ballistic {Transport} {Exceeding} 28 $\mu$m in {CVD} {Grown}
  {Graphene}},\ }\href {https://doi.org/10.1021/acs.nanolett.5b04840}
  {\bibfield  {journal} {\bibinfo  {journal} {Nano Letters}\ }\textbf {\bibinfo
  {volume} {16}},\ \bibinfo {pages} {1387} (\bibinfo {year}
  {2016})}\BibitemShut {NoStop}%
\bibitem [{\citenamefont {Hasan}\ and\ \citenamefont
  {Kane}(2010)}]{hasan_topological_2010}%
  \BibitemOpen
  \bibfield  {author} {\bibinfo {author} {\bibfnamefont {M.~Z.}\ \bibnamefont
  {Hasan}}\ and\ \bibinfo {author} {\bibfnamefont {C.~L.}\ \bibnamefont
  {Kane}},\ }\bibfield  {title} {\bibinfo {title} {Topological insulators},\
  }\href {https://doi.org/10.1103/RevModPhys.82.3045} {\bibfield  {journal}
  {\bibinfo  {journal} {Reviews of Modern Physics}\ }\textbf {\bibinfo {volume}
  {82}},\ \bibinfo {pages} {3045} (\bibinfo {year} {2010})}\BibitemShut
  {NoStop}%
\bibitem [{\citenamefont {Qi}\ and\ \citenamefont
  {Zhang}(2011)}]{qi_topological_2011}%
  \BibitemOpen
  \bibfield  {author} {\bibinfo {author} {\bibfnamefont {X.-L.}\ \bibnamefont
  {Qi}}\ and\ \bibinfo {author} {\bibfnamefont {S.-C.}\ \bibnamefont {Zhang}},\
  }\bibfield  {title} {\bibinfo {title} {Topological insulators and
  superconductors},\ }\href {https://doi.org/10.1103/RevModPhys.83.1057}
  {\bibfield  {journal} {\bibinfo  {journal} {Reviews of Modern Physics}\
  }\textbf {\bibinfo {volume} {83}},\ \bibinfo {pages} {1057} (\bibinfo {year}
  {2011})}\BibitemShut {NoStop}%
\bibitem [{\citenamefont {Armitage}\ \emph {et~al.}(2018)\citenamefont
  {Armitage}, \citenamefont {Mele},\ and\ \citenamefont
  {Vishwanath}}]{armitage_weyl_2018}%
  \BibitemOpen
  \bibfield  {author} {\bibinfo {author} {\bibfnamefont {N.~P.}\ \bibnamefont
  {Armitage}}, \bibinfo {author} {\bibfnamefont {E.~J.}\ \bibnamefont {Mele}},\
  and\ \bibinfo {author} {\bibfnamefont {A.}~\bibnamefont {Vishwanath}},\
  }\bibfield  {title} {\bibinfo {title} {Weyl and {Dirac} semimetals in
  three-dimensional solids},\ }\href
  {https://doi.org/10.1103/RevModPhys.90.015001} {\bibfield  {journal}
  {\bibinfo  {journal} {Reviews of Modern Physics}\ }\textbf {\bibinfo {volume}
  {90}},\ \bibinfo {pages} {015001} (\bibinfo {year} {2018})}\BibitemShut
  {NoStop}%
\bibitem [{\citenamefont {Liang}\ \emph {et~al.}(2015)\citenamefont {Liang},
  \citenamefont {Gibson}, \citenamefont {Ali}, \citenamefont {Liu},
  \citenamefont {Cava},\ and\ \citenamefont {Ong}}]{liang_ultrahigh_2015}%
  \BibitemOpen
  \bibfield  {author} {\bibinfo {author} {\bibfnamefont {T.}~\bibnamefont
  {Liang}}, \bibinfo {author} {\bibfnamefont {Q.}~\bibnamefont {Gibson}},
  \bibinfo {author} {\bibfnamefont {M.~N.}\ \bibnamefont {Ali}}, \bibinfo
  {author} {\bibfnamefont {M.}~\bibnamefont {Liu}}, \bibinfo {author}
  {\bibfnamefont {R.~J.}\ \bibnamefont {Cava}},\ and\ \bibinfo {author}
  {\bibfnamefont {N.~P.}\ \bibnamefont {Ong}},\ }\bibfield  {title}
  {{\selectlanguage {english}\bibinfo {title} {Ultrahigh mobility and giant
  magnetoresistance in the {Dirac} semimetal
  {Cd}$_{\textrm{3}}${As}$_{\textrm{2}}$}},\ }\href
  {https://doi.org/10.1038/nmat4143} {\bibfield  {journal} {\bibinfo  {journal}
  {Nature Materials}\ }\textbf {\bibinfo {volume} {14}},\ \bibinfo {pages}
  {280} (\bibinfo {year} {2015})}\BibitemShut {NoStop}%
\bibitem [{\citenamefont {Shekhar}\ \emph {et~al.}(2015)\citenamefont
  {Shekhar}, \citenamefont {Nayak}, \citenamefont {Sun}, \citenamefont
  {Schmidt}, \citenamefont {Nicklas}, \citenamefont {Leermakers}, \citenamefont
  {Zeitler}, \citenamefont {Skourski}, \citenamefont {Wosnitza}, \citenamefont
  {Liu}, \citenamefont {Chen}, \citenamefont {Schnelle}, \citenamefont
  {Borrmann}, \citenamefont {Grin}, \citenamefont {Felser},\ and\ \citenamefont
  {Yan}}]{shekhar_extremely_2015}%
  \BibitemOpen
  \bibfield  {author} {\bibinfo {author} {\bibfnamefont {C.}~\bibnamefont
  {Shekhar}}, \bibinfo {author} {\bibfnamefont {A.~K.}\ \bibnamefont {Nayak}},
  \bibinfo {author} {\bibfnamefont {Y.}~\bibnamefont {Sun}}, \bibinfo {author}
  {\bibfnamefont {M.}~\bibnamefont {Schmidt}}, \bibinfo {author} {\bibfnamefont
  {M.}~\bibnamefont {Nicklas}}, \bibinfo {author} {\bibfnamefont
  {I.}~\bibnamefont {Leermakers}}, \bibinfo {author} {\bibfnamefont
  {U.}~\bibnamefont {Zeitler}}, \bibinfo {author} {\bibfnamefont
  {Y.}~\bibnamefont {Skourski}}, \bibinfo {author} {\bibfnamefont
  {J.}~\bibnamefont {Wosnitza}}, \bibinfo {author} {\bibfnamefont
  {Z.}~\bibnamefont {Liu}}, \bibinfo {author} {\bibfnamefont {Y.}~\bibnamefont
  {Chen}}, \bibinfo {author} {\bibfnamefont {W.}~\bibnamefont {Schnelle}},
  \bibinfo {author} {\bibfnamefont {H.}~\bibnamefont {Borrmann}}, \bibinfo
  {author} {\bibfnamefont {Y.}~\bibnamefont {Grin}}, \bibinfo {author}
  {\bibfnamefont {C.}~\bibnamefont {Felser}},\ and\ \bibinfo {author}
  {\bibfnamefont {B.}~\bibnamefont {Yan}},\ }\bibfield  {title}
  {{\selectlanguage {english}\bibinfo {title} {Extremely large
  magnetoresistance and ultrahigh mobility in the topological {Weyl} semimetal
  candidate {NbP}}},\ }\href {https://doi.org/10.1038/nphys3372} {\bibfield
  {journal} {\bibinfo  {journal} {Nature Physics}\ }\textbf {\bibinfo {volume}
  {11}},\ \bibinfo {pages} {645} (\bibinfo {year} {2015})}\BibitemShut
  {NoStop}%
\bibitem [{\citenamefont {Kumar}\ \emph {et~al.}(2017)\citenamefont {Kumar},
  \citenamefont {Sun}, \citenamefont {Xu}, \citenamefont {Manna}, \citenamefont
  {Yao}, \citenamefont {S{\"u}ss}, \citenamefont {Leermakers}, \citenamefont
  {Young}, \citenamefont {F{\"o}rster}, \citenamefont {Schmidt}, \citenamefont
  {Borrmann}, \citenamefont {Yan}, \citenamefont {Zeitler}, \citenamefont
  {Shi}, \citenamefont {Felser},\ and\ \citenamefont
  {Shekhar}}]{kumar_extremely_2017}%
  \BibitemOpen
  \bibfield  {author} {\bibinfo {author} {\bibfnamefont {N.}~\bibnamefont
  {Kumar}}, \bibinfo {author} {\bibfnamefont {Y.}~\bibnamefont {Sun}}, \bibinfo
  {author} {\bibfnamefont {N.}~\bibnamefont {Xu}}, \bibinfo {author}
  {\bibfnamefont {K.}~\bibnamefont {Manna}}, \bibinfo {author} {\bibfnamefont
  {M.}~\bibnamefont {Yao}}, \bibinfo {author} {\bibfnamefont {V.}~\bibnamefont
  {S{\"u}ss}}, \bibinfo {author} {\bibfnamefont {I.}~\bibnamefont
  {Leermakers}}, \bibinfo {author} {\bibfnamefont {O.}~\bibnamefont {Young}},
  \bibinfo {author} {\bibfnamefont {T.}~\bibnamefont {F{\"o}rster}}, \bibinfo
  {author} {\bibfnamefont {M.}~\bibnamefont {Schmidt}}, \bibinfo {author}
  {\bibfnamefont {H.}~\bibnamefont {Borrmann}}, \bibinfo {author}
  {\bibfnamefont {B.}~\bibnamefont {Yan}}, \bibinfo {author} {\bibfnamefont
  {U.}~\bibnamefont {Zeitler}}, \bibinfo {author} {\bibfnamefont
  {M.}~\bibnamefont {Shi}}, \bibinfo {author} {\bibfnamefont {C.}~\bibnamefont
  {Felser}},\ and\ \bibinfo {author} {\bibfnamefont {C.}~\bibnamefont
  {Shekhar}},\ }\bibfield  {title} {{\selectlanguage {english}\bibinfo {title}
  {Extremely high magnetoresistance and conductivity in the type-{II} {Weyl}
  semimetals {WP} 2 and {MoP} 2}},\ }\href
  {https://doi.org/10.1038/s41467-017-01758-z} {\bibfield  {journal} {\bibinfo
  {journal} {Nature Communications}\ }\textbf {\bibinfo {volume} {8}},\
  \bibinfo {pages} {1} (\bibinfo {year} {2017})}\BibitemShut {NoStop}%
\bibitem [{\citenamefont {Zhang}\ \emph {et~al.}(2019)\citenamefont {Zhang},
  \citenamefont {Ni}, \citenamefont {Zhang}, \citenamefont {Yuan},
  \citenamefont {Liu}, \citenamefont {Zou}, \citenamefont {Liao}, \citenamefont
  {Du}, \citenamefont {Narayan}, \citenamefont {Zhang}, \citenamefont {Gu},
  \citenamefont {Zhu}, \citenamefont {Pi}, \citenamefont {Sanvito},
  \citenamefont {Han}, \citenamefont {Zou}, \citenamefont {Shi}, \citenamefont
  {Wan}, \citenamefont {Savrasov},\ and\ \citenamefont
  {Xiu}}]{zhang_ultrahigh_2019}%
  \BibitemOpen
  \bibfield  {author} {\bibinfo {author} {\bibfnamefont {C.}~\bibnamefont
  {Zhang}}, \bibinfo {author} {\bibfnamefont {Z.}~\bibnamefont {Ni}}, \bibinfo
  {author} {\bibfnamefont {J.}~\bibnamefont {Zhang}}, \bibinfo {author}
  {\bibfnamefont {X.}~\bibnamefont {Yuan}}, \bibinfo {author} {\bibfnamefont
  {Y.}~\bibnamefont {Liu}}, \bibinfo {author} {\bibfnamefont {Y.}~\bibnamefont
  {Zou}}, \bibinfo {author} {\bibfnamefont {Z.}~\bibnamefont {Liao}}, \bibinfo
  {author} {\bibfnamefont {Y.}~\bibnamefont {Du}}, \bibinfo {author}
  {\bibfnamefont {A.}~\bibnamefont {Narayan}}, \bibinfo {author} {\bibfnamefont
  {H.}~\bibnamefont {Zhang}}, \bibinfo {author} {\bibfnamefont
  {T.}~\bibnamefont {Gu}}, \bibinfo {author} {\bibfnamefont {X.}~\bibnamefont
  {Zhu}}, \bibinfo {author} {\bibfnamefont {L.}~\bibnamefont {Pi}}, \bibinfo
  {author} {\bibfnamefont {S.}~\bibnamefont {Sanvito}}, \bibinfo {author}
  {\bibfnamefont {X.}~\bibnamefont {Han}}, \bibinfo {author} {\bibfnamefont
  {J.}~\bibnamefont {Zou}}, \bibinfo {author} {\bibfnamefont {Y.}~\bibnamefont
  {Shi}}, \bibinfo {author} {\bibfnamefont {X.}~\bibnamefont {Wan}}, \bibinfo
  {author} {\bibfnamefont {S.~Y.}\ \bibnamefont {Savrasov}},\ and\ \bibinfo
  {author} {\bibfnamefont {F.}~\bibnamefont {Xiu}},\ }\bibfield  {title}
  {{\selectlanguage {english}\bibinfo {title} {Ultrahigh conductivity in {Weyl}
  semimetal {NbAs} nanobelts}},\ }\href
  {https://doi.org/10.1038/s41563-019-0320-9} {\bibfield  {journal} {\bibinfo
  {journal} {Nature Materials}\ }\textbf {\bibinfo {volume} {18}},\ \bibinfo
  {pages} {482} (\bibinfo {year} {2019})}\BibitemShut {NoStop}%
\bibitem [{\citenamefont {Shannon}\ \emph {et~al.}(1971)\citenamefont
  {Shannon}, \citenamefont {Prewitt},\ and\ \citenamefont
  {Rogers}}]{shannon_chemistry_1971}%
  \BibitemOpen
  \bibfield  {author} {\bibinfo {author} {\bibfnamefont {R.~D.}\ \bibnamefont
  {Shannon}}, \bibinfo {author} {\bibfnamefont {C.~T.}\ \bibnamefont
  {Prewitt}},\ and\ \bibinfo {author} {\bibfnamefont {D.~B.}\ \bibnamefont
  {Rogers}},\ }\bibfield  {title} {\bibinfo {title} {Chemistry of noble metal
  oxides. {II}. {Crystal} structures of platinum cobalt dioxide, palladium
  cobalt dioxide, copper iron dioxide, and silver iron dioxide},\ }\href
  {https://doi.org/10.1021/ic50098a012} {\bibfield  {journal} {\bibinfo
  {journal} {Inorganic Chemistry}\ }\textbf {\bibinfo {volume} {10}},\ \bibinfo
  {pages} {719} (\bibinfo {year} {1971})}\BibitemShut {NoStop}%
\bibitem [{\citenamefont {Mackenzie}(2017)}]{mackenzie_properties_2017}%
  \BibitemOpen
  \bibfield  {author} {\bibinfo {author} {\bibfnamefont {A.~P.}\ \bibnamefont
  {Mackenzie}},\ }\bibfield  {title} {{\selectlanguage {english}\bibinfo
  {title} {The properties of ultrapure delafossite metals}},\ }\href
  {https://doi.org/10.1088/1361-6633/aa50e5} {\bibfield  {journal} {\bibinfo
  {journal} {Reports on Progress in Physics}\ }\textbf {\bibinfo {volume}
  {80}},\ \bibinfo {pages} {032501} (\bibinfo {year} {2017})}\BibitemShut
  {NoStop}%
\bibitem [{\citenamefont {Takatsu}\ \emph {et~al.}(2007)\citenamefont
  {Takatsu}, \citenamefont {Yonezawa}, \citenamefont {Mouri}, \citenamefont
  {Nakatsuji}, \citenamefont {Tanaka},\ and\ \citenamefont
  {Maeno}}]{takatsu_roles_2007}%
  \BibitemOpen
  \bibfield  {author} {\bibinfo {author} {\bibfnamefont {H.}~\bibnamefont
  {Takatsu}}, \bibinfo {author} {\bibfnamefont {S.}~\bibnamefont {Yonezawa}},
  \bibinfo {author} {\bibfnamefont {S.}~\bibnamefont {Mouri}}, \bibinfo
  {author} {\bibfnamefont {S.}~\bibnamefont {Nakatsuji}}, \bibinfo {author}
  {\bibfnamefont {K.}~\bibnamefont {Tanaka}},\ and\ \bibinfo {author}
  {\bibfnamefont {Y.}~\bibnamefont {Maeno}},\ }\bibfield  {title}
  {{\selectlanguage {english}\bibinfo {title} {Roles of {High}-{Frequency}
  {Optical} {Phonons} in the {Physical} {Properties} of the {Conductive}
  {Delafossite} {PdCoO}$_{\textrm{2}}$}},\ }\href
  {https://doi.org/10.1143/JPSJ.76.104701} {\bibfield  {journal} {\bibinfo
  {journal} {Journal of the Physical Society of Japan}\ }\textbf {\bibinfo
  {volume} {76}},\ \bibinfo {pages} {104701} (\bibinfo {year}
  {2007})}\BibitemShut {NoStop}%
\bibitem [{\citenamefont {Takatsu}\ \emph {et~al.}(2010)\citenamefont
  {Takatsu}, \citenamefont {Yonezawa}, \citenamefont {Michioka}, \citenamefont
  {Yoshimura},\ and\ \citenamefont {Maeno}}]{takatsu_anisotropy_2010}%
  \BibitemOpen
  \bibfield  {author} {\bibinfo {author} {\bibfnamefont {H.}~\bibnamefont
  {Takatsu}}, \bibinfo {author} {\bibfnamefont {S.}~\bibnamefont {Yonezawa}},
  \bibinfo {author} {\bibfnamefont {C.}~\bibnamefont {Michioka}}, \bibinfo
  {author} {\bibfnamefont {K.}~\bibnamefont {Yoshimura}},\ and\ \bibinfo
  {author} {\bibfnamefont {Y.}~\bibnamefont {Maeno}},\ }\bibfield  {title}
  {{\selectlanguage {english}\bibinfo {title} {Anisotropy in the magnetization
  and resistivity of the metallic triangular-lattice magnet
  {PdCrO}$_{\textrm{2}}$}},\ }\href
  {https://doi.org/10.1088/1742-6596/200/1/012198} {\bibfield  {journal}
  {\bibinfo  {journal} {Journal of Physics: Conference Series}\ }\textbf
  {\bibinfo {volume} {200}},\ \bibinfo {pages} {012198} (\bibinfo {year}
  {2010})}\BibitemShut {NoStop}%
\bibitem [{\citenamefont {Kushwaha}\ \emph {et~al.}(2015)\citenamefont
  {Kushwaha}, \citenamefont {Sunko}, \citenamefont {Moll}, \citenamefont
  {Bawden}, \citenamefont {Riley}, \citenamefont {Nandi}, \citenamefont
  {Rosner}, \citenamefont {Schmidt}, \citenamefont {Arnold}, \citenamefont
  {Hassinger}, \citenamefont {Kim}, \citenamefont {Hoesch}, \citenamefont
  {Mackenzie},\ and\ \citenamefont {King}}]{kushwaha_nearly_2015}%
  \BibitemOpen
  \bibfield  {author} {\bibinfo {author} {\bibfnamefont {P.}~\bibnamefont
  {Kushwaha}}, \bibinfo {author} {\bibfnamefont {V.}~\bibnamefont {Sunko}},
  \bibinfo {author} {\bibfnamefont {P.~J.~W.}\ \bibnamefont {Moll}}, \bibinfo
  {author} {\bibfnamefont {L.}~\bibnamefont {Bawden}}, \bibinfo {author}
  {\bibfnamefont {J.~M.}\ \bibnamefont {Riley}}, \bibinfo {author}
  {\bibfnamefont {N.}~\bibnamefont {Nandi}}, \bibinfo {author} {\bibfnamefont
  {H.}~\bibnamefont {Rosner}}, \bibinfo {author} {\bibfnamefont {M.~P.}\
  \bibnamefont {Schmidt}}, \bibinfo {author} {\bibfnamefont {F.}~\bibnamefont
  {Arnold}}, \bibinfo {author} {\bibfnamefont {E.}~\bibnamefont {Hassinger}},
  \bibinfo {author} {\bibfnamefont {T.~K.}\ \bibnamefont {Kim}}, \bibinfo
  {author} {\bibfnamefont {M.}~\bibnamefont {Hoesch}}, \bibinfo {author}
  {\bibfnamefont {A.~P.}\ \bibnamefont {Mackenzie}},\ and\ \bibinfo {author}
  {\bibfnamefont {P.~D.~C.}\ \bibnamefont {King}},\ }\bibfield  {title}
  {{\selectlanguage {english}\bibinfo {title} {Nearly free electrons in a
  5\textit{d} delafossite oxide metal}},\ }\href
  {https://doi.org/10.1126/sciadv.1500692} {\bibfield  {journal} {\bibinfo
  {journal} {Science Advances}\ }\textbf {\bibinfo {volume} {1}},\ \bibinfo
  {pages} {e1500692} (\bibinfo {year} {2015})}\BibitemShut {NoStop}%
\bibitem [{\citenamefont {Hicks}\ \emph {et~al.}(2012)\citenamefont {Hicks},
  \citenamefont {Gibbs}, \citenamefont {Mackenzie}, \citenamefont {Takatsu},
  \citenamefont {Maeno},\ and\ \citenamefont {Yelland}}]{hicks_quantum_2012}%
  \BibitemOpen
  \bibfield  {author} {\bibinfo {author} {\bibfnamefont {C.~W.}\ \bibnamefont
  {Hicks}}, \bibinfo {author} {\bibfnamefont {A.~S.}\ \bibnamefont {Gibbs}},
  \bibinfo {author} {\bibfnamefont {A.~P.}\ \bibnamefont {Mackenzie}}, \bibinfo
  {author} {\bibfnamefont {H.}~\bibnamefont {Takatsu}}, \bibinfo {author}
  {\bibfnamefont {Y.}~\bibnamefont {Maeno}},\ and\ \bibinfo {author}
  {\bibfnamefont {E.~A.}\ \bibnamefont {Yelland}},\ }\bibfield  {title}
  {\bibinfo {title} {Quantum oscillations and high carrier mobility in the
  delafossite {PdCoO}$_{\textrm{2}}$},\ }\href
  {https://doi.org/10.1103/PhysRevLett.109.116401} {\bibfield  {journal}
  {\bibinfo  {journal} {Physical Review Letters}\ }\textbf {\bibinfo {volume}
  {109}},\ \bibinfo {pages} {116401} (\bibinfo {year} {2012})}\BibitemShut
  {NoStop}%
\bibitem [{\citenamefont {Takatsu}\ \emph {et~al.}(2013)\citenamefont
  {Takatsu}, \citenamefont {Ishikawa}, \citenamefont {Yonezawa}, \citenamefont
  {Yoshino}, \citenamefont {Shishidou}, \citenamefont {Oguchi}, \citenamefont
  {Murata},\ and\ \citenamefont {Maeno}}]{takatsu_extremely_2013}%
  \BibitemOpen
  \bibfield  {author} {\bibinfo {author} {\bibfnamefont {H.}~\bibnamefont
  {Takatsu}}, \bibinfo {author} {\bibfnamefont {J.~J.}\ \bibnamefont
  {Ishikawa}}, \bibinfo {author} {\bibfnamefont {S.}~\bibnamefont {Yonezawa}},
  \bibinfo {author} {\bibfnamefont {H.}~\bibnamefont {Yoshino}}, \bibinfo
  {author} {\bibfnamefont {T.}~\bibnamefont {Shishidou}}, \bibinfo {author}
  {\bibfnamefont {T.}~\bibnamefont {Oguchi}}, \bibinfo {author} {\bibfnamefont
  {K.}~\bibnamefont {Murata}},\ and\ \bibinfo {author} {\bibfnamefont
  {Y.}~\bibnamefont {Maeno}},\ }\bibfield  {title} {\bibinfo {title} {Extremely
  {Large} {Magnetoresistance} in the {Nonmagnetic} {Metal}
  {PdCoO}$_{\textrm{2}}$},\ }\href
  {https://doi.org/10.1103/PhysRevLett.111.056601} {\bibfield  {journal}
  {\bibinfo  {journal} {Physical Review Letters}\ }\textbf {\bibinfo {volume}
  {111}},\ \bibinfo {pages} {056601} (\bibinfo {year} {2013})}\BibitemShut
  {NoStop}%
\bibitem [{\citenamefont {Daou}\ \emph {et~al.}(2015)\citenamefont {Daou},
  \citenamefont {Fr{\'e}sard}, \citenamefont {H{\'e}bert},\ and\ \citenamefont
  {Maignan}}]{daou_large_2015}%
  \BibitemOpen
  \bibfield  {author} {\bibinfo {author} {\bibfnamefont {R.}~\bibnamefont
  {Daou}}, \bibinfo {author} {\bibfnamefont {R.}~\bibnamefont {Fr{\'e}sard}},
  \bibinfo {author} {\bibfnamefont {S.}~\bibnamefont {H{\'e}bert}},\ and\
  \bibinfo {author} {\bibfnamefont {A.}~\bibnamefont {Maignan}},\ }\bibfield
  {title} {\bibinfo {title} {Large anisotropic thermal conductivity of the
  intrinsically two-dimensional metallic oxide {PdCoO}$_{\textrm{2}}$},\ }\href
  {https://doi.org/10.1103/PhysRevB.91.041113} {\bibfield  {journal} {\bibinfo
  {journal} {Phys. Rev. B}\ }\textbf {\bibinfo {volume} {91}},\ \bibinfo
  {pages} {041113} (\bibinfo {year} {2015})}\BibitemShut {NoStop}%
\bibitem [{\citenamefont {Moll}\ \emph {et~al.}(2016)\citenamefont {Moll},
  \citenamefont {Kushwaha}, \citenamefont {Nandi}, \citenamefont {Schmidt},\
  and\ \citenamefont {Mackenzie}}]{moll_evidence_2016}%
  \BibitemOpen
  \bibfield  {author} {\bibinfo {author} {\bibfnamefont {P.~J.~W.}\
  \bibnamefont {Moll}}, \bibinfo {author} {\bibfnamefont {P.}~\bibnamefont
  {Kushwaha}}, \bibinfo {author} {\bibfnamefont {N.}~\bibnamefont {Nandi}},
  \bibinfo {author} {\bibfnamefont {B.}~\bibnamefont {Schmidt}},\ and\ \bibinfo
  {author} {\bibfnamefont {A.~P.}\ \bibnamefont {Mackenzie}},\ }\bibfield
  {title} {{\selectlanguage {english}\bibinfo {title} {Evidence for
  hydrodynamic electron flow in {PdCoO}$_{\textrm{2}}$}},\ }\href
  {https://doi.org/10.1126/science.aac8385} {\bibfield  {journal} {\bibinfo
  {journal} {Science}\ }\textbf {\bibinfo {volume} {351}},\ \bibinfo {pages}
  {1061} (\bibinfo {year} {2016})}\BibitemShut {NoStop}%
\bibitem [{\citenamefont {Kikugawa}\ \emph {et~al.}(2016)\citenamefont
  {Kikugawa}, \citenamefont {Goswami}, \citenamefont {Kiswandhi}, \citenamefont
  {Choi}, \citenamefont {Graf}, \citenamefont {Baumbach}, \citenamefont
  {Brooks}, \citenamefont {Sugii}, \citenamefont {Iida}, \citenamefont
  {Nishio}, \citenamefont {Uji}, \citenamefont {Terashima}, \citenamefont
  {Rourke}, \citenamefont {Hussey}, \citenamefont {Takatsu}, \citenamefont
  {Yonezawa}, \citenamefont {Maeno},\ and\ \citenamefont
  {Balicas}}]{kikugawa_interplanar_2016}%
  \BibitemOpen
  \bibfield  {author} {\bibinfo {author} {\bibfnamefont {N.}~\bibnamefont
  {Kikugawa}}, \bibinfo {author} {\bibfnamefont {P.}~\bibnamefont {Goswami}},
  \bibinfo {author} {\bibfnamefont {A.}~\bibnamefont {Kiswandhi}}, \bibinfo
  {author} {\bibfnamefont {E.~S.}\ \bibnamefont {Choi}}, \bibinfo {author}
  {\bibfnamefont {D.}~\bibnamefont {Graf}}, \bibinfo {author} {\bibfnamefont
  {R.~E.}\ \bibnamefont {Baumbach}}, \bibinfo {author} {\bibfnamefont {J.~S.}\
  \bibnamefont {Brooks}}, \bibinfo {author} {\bibfnamefont {K.}~\bibnamefont
  {Sugii}}, \bibinfo {author} {\bibfnamefont {Y.}~\bibnamefont {Iida}},
  \bibinfo {author} {\bibfnamefont {M.}~\bibnamefont {Nishio}}, \bibinfo
  {author} {\bibfnamefont {S.}~\bibnamefont {Uji}}, \bibinfo {author}
  {\bibfnamefont {T.}~\bibnamefont {Terashima}}, \bibinfo {author}
  {\bibfnamefont {P.~M.~C.}\ \bibnamefont {Rourke}}, \bibinfo {author}
  {\bibfnamefont {N.~E.}\ \bibnamefont {Hussey}}, \bibinfo {author}
  {\bibfnamefont {H.}~\bibnamefont {Takatsu}}, \bibinfo {author} {\bibfnamefont
  {S.}~\bibnamefont {Yonezawa}}, \bibinfo {author} {\bibfnamefont
  {Y.}~\bibnamefont {Maeno}},\ and\ \bibinfo {author} {\bibfnamefont
  {L.}~\bibnamefont {Balicas}},\ }\bibfield  {title} {{\selectlanguage
  {english}\bibinfo {title} {Interplanar coupling-dependent magnetoresistivity
  in high-purity layered metals}},\ }\href
  {https://doi.org/10.1038/ncomms10903} {\bibfield  {journal} {\bibinfo
  {journal} {Nature Communications}\ }\textbf {\bibinfo {volume} {7}},\
  \bibinfo {pages} {10903} (\bibinfo {year} {2016})}\BibitemShut {NoStop}%
\bibitem [{\citenamefont {Nandi}\ \emph {et~al.}(2018)\citenamefont {Nandi},
  \citenamefont {Scaffidi}, \citenamefont {Kushwaha}, \citenamefont {Khim},
  \citenamefont {Barber}, \citenamefont {Sunko}, \citenamefont {Mazzola},
  \citenamefont {King}, \citenamefont {Rosner}, \citenamefont {Moll},
  \citenamefont {K{\"o}nig}, \citenamefont {Moore}, \citenamefont {Hartnoll},\
  and\ \citenamefont {Mackenzie}}]{nandi_unconventional_2018}%
  \BibitemOpen
  \bibfield  {author} {\bibinfo {author} {\bibfnamefont {N.}~\bibnamefont
  {Nandi}}, \bibinfo {author} {\bibfnamefont {T.}~\bibnamefont {Scaffidi}},
  \bibinfo {author} {\bibfnamefont {P.}~\bibnamefont {Kushwaha}}, \bibinfo
  {author} {\bibfnamefont {S.}~\bibnamefont {Khim}}, \bibinfo {author}
  {\bibfnamefont {M.~E.}\ \bibnamefont {Barber}}, \bibinfo {author}
  {\bibfnamefont {V.}~\bibnamefont {Sunko}}, \bibinfo {author} {\bibfnamefont
  {F.}~\bibnamefont {Mazzola}}, \bibinfo {author} {\bibfnamefont {P.~D.~C.}\
  \bibnamefont {King}}, \bibinfo {author} {\bibfnamefont {H.}~\bibnamefont
  {Rosner}}, \bibinfo {author} {\bibfnamefont {P.~J.~W.}\ \bibnamefont {Moll}},
  \bibinfo {author} {\bibfnamefont {M.}~\bibnamefont {K{\"o}nig}}, \bibinfo
  {author} {\bibfnamefont {J.~E.}\ \bibnamefont {Moore}}, \bibinfo {author}
  {\bibfnamefont {S.}~\bibnamefont {Hartnoll}},\ and\ \bibinfo {author}
  {\bibfnamefont {A.~P.}\ \bibnamefont {Mackenzie}},\ }\bibfield  {title}
  {{\selectlanguage {english}\bibinfo {title} {Unconventional magneto-transport
  in ultrapure {PdCoO}$_{\textrm{2}}$ and {PtCoO}$_{\textrm{2}}$}},\ }\href
  {https://doi.org/10.1038/s41535-018-0138-8} {\bibfield  {journal} {\bibinfo
  {journal} {npj Quantum Materials}\ }\textbf {\bibinfo {volume} {3}},\
  \bibinfo {pages} {66} (\bibinfo {year} {2018})}\BibitemShut {NoStop}%
\bibitem [{\citenamefont {Putzke}\ \emph {et~al.}(2019)\citenamefont {Putzke},
  \citenamefont {Bachmann}, \citenamefont {McGuinness}, \citenamefont
  {Zhakina}, \citenamefont {Oka}, \citenamefont {Moessner}, \citenamefont
  {K{\"o}nig}, \citenamefont {Khim}, \citenamefont {Mackenzie},\ and\
  \citenamefont {Moll}}]{putzke_h/e_2019}%
  \BibitemOpen
  \bibfield  {author} {\bibinfo {author} {\bibfnamefont {C.}~\bibnamefont
  {Putzke}}, \bibinfo {author} {\bibfnamefont {M.~D.}\ \bibnamefont
  {Bachmann}}, \bibinfo {author} {\bibfnamefont {P.}~\bibnamefont
  {McGuinness}}, \bibinfo {author} {\bibfnamefont {E.}~\bibnamefont {Zhakina}},
  \bibinfo {author} {\bibfnamefont {T.}~\bibnamefont {Oka}}, \bibinfo {author}
  {\bibfnamefont {R.}~\bibnamefont {Moessner}}, \bibinfo {author}
  {\bibfnamefont {M.}~\bibnamefont {K{\"o}nig}}, \bibinfo {author}
  {\bibfnamefont {S.}~\bibnamefont {Khim}}, \bibinfo {author} {\bibfnamefont
  {A.~P.}\ \bibnamefont {Mackenzie}},\ and\ \bibinfo {author} {\bibfnamefont
  {P.~J.~W.}\ \bibnamefont {Moll}},\ }\bibfield  {title} {\bibinfo {title} {h/e
  {Oscillations} in {Interlayer} {Transport} of {Delafossites}},\ }\href
  {http://arxiv.org/abs/1902.07331} {\bibfield  {journal} {\bibinfo  {journal}
  {arXiv:1902.07331}\ } (\bibinfo {year} {2019})}\BibitemShut {NoStop}%
\bibitem [{\citenamefont {Usui}\ \emph {et~al.}(2019)\citenamefont {Usui},
  \citenamefont {Ochi}, \citenamefont {Kitamura}, \citenamefont {Oka},
  \citenamefont {Ogura}, \citenamefont {Rosner}, \citenamefont {Haverkort},
  \citenamefont {Sunko}, \citenamefont {King}, \citenamefont {Mackenzie},\ and\
  \citenamefont {Kuroki}}]{usui_hidden_2019}%
  \BibitemOpen
  \bibfield  {author} {\bibinfo {author} {\bibfnamefont {H.}~\bibnamefont
  {Usui}}, \bibinfo {author} {\bibfnamefont {M.}~\bibnamefont {Ochi}}, \bibinfo
  {author} {\bibfnamefont {S.}~\bibnamefont {Kitamura}}, \bibinfo {author}
  {\bibfnamefont {T.}~\bibnamefont {Oka}}, \bibinfo {author} {\bibfnamefont
  {D.}~\bibnamefont {Ogura}}, \bibinfo {author} {\bibfnamefont
  {H.}~\bibnamefont {Rosner}}, \bibinfo {author} {\bibfnamefont {M.~W.}\
  \bibnamefont {Haverkort}}, \bibinfo {author} {\bibfnamefont {V.}~\bibnamefont
  {Sunko}}, \bibinfo {author} {\bibfnamefont {P.~D.~C.}\ \bibnamefont {King}},
  \bibinfo {author} {\bibfnamefont {A.~P.}\ \bibnamefont {Mackenzie}},\ and\
  \bibinfo {author} {\bibfnamefont {K.}~\bibnamefont {Kuroki}},\ }\bibfield
  {title} {\bibinfo {title} {Hidden kagome-lattice picture and origin of high
  conductivity in delafossite {PtCoO}$_{\textrm{2}}$},\ }\href
  {https://doi.org/10.1103/PhysRevMaterials.3.045002} {\bibfield  {journal}
  {\bibinfo  {journal} {Physical Review Materials}\ }\textbf {\bibinfo {volume}
  {3}},\ \bibinfo {pages} {045002} (\bibinfo {year} {2019})}\BibitemShut
  {NoStop}%
\bibitem [{\citenamefont {Kushwaha}\ \emph {et~al.}(2017)\citenamefont
  {Kushwaha}, \citenamefont {Borrmann}, \citenamefont {Khim}, \citenamefont
  {Rosner}, \citenamefont {Moll}, \citenamefont {Sokolov}, \citenamefont
  {Sunko}, \citenamefont {Grin},\ and\ \citenamefont
  {Mackenzie}}]{kushwaha_single_2017}%
  \BibitemOpen
  \bibfield  {author} {\bibinfo {author} {\bibfnamefont {P.}~\bibnamefont
  {Kushwaha}}, \bibinfo {author} {\bibfnamefont {H.}~\bibnamefont {Borrmann}},
  \bibinfo {author} {\bibfnamefont {S.}~\bibnamefont {Khim}}, \bibinfo {author}
  {\bibfnamefont {H.}~\bibnamefont {Rosner}}, \bibinfo {author} {\bibfnamefont
  {P.~J.~W.}\ \bibnamefont {Moll}}, \bibinfo {author} {\bibfnamefont {D.~A.}\
  \bibnamefont {Sokolov}}, \bibinfo {author} {\bibfnamefont {V.}~\bibnamefont
  {Sunko}}, \bibinfo {author} {\bibfnamefont {Y.}~\bibnamefont {Grin}},\ and\
  \bibinfo {author} {\bibfnamefont {A.~P.}\ \bibnamefont {Mackenzie}},\
  }\bibfield  {title} {\bibinfo {title} {Single {Crystal} {Growth},
  {Structure}, and {Electronic} {Properties} of {Metallic} {Delafossite}
  {PdRhO}$_{\textrm{2}}$},\ }\href {https://doi.org/10.1021/acs.cgd.7b00418}
  {\bibfield  {journal} {\bibinfo  {journal} {Crystal Growth \& Design}\
  }\textbf {\bibinfo {volume} {17}},\ \bibinfo {pages} {4144} (\bibinfo {year}
  {2017})}\BibitemShut {NoStop}%
\bibitem [{\citenamefont {Berger}\ \emph {et~al.}(2019)\citenamefont {Berger},
  \citenamefont {Coursey}, \citenamefont {Zucker},\ and\ \citenamefont
  {Chang}}]{berger_star_2019}%
  \BibitemOpen
  \bibfield  {author} {\bibinfo {author} {\bibfnamefont {M.}~\bibnamefont
  {Berger}}, \bibinfo {author} {\bibfnamefont {J.}~\bibnamefont {Coursey}},
  \bibinfo {author} {\bibfnamefont {M.}~\bibnamefont {Zucker}},\ and\ \bibinfo
  {author} {\bibfnamefont {J.}~\bibnamefont {Chang}},\ }\href
  {http://physics.nist.gov/Star} {\bibinfo {title} {{STAR}, {PSTAR}, and
  {ASTAR}: {Computer} {Programs} for {Calculating} {Stopping}-{Power} and
  {Range} {Tables} for {Electrons}, {Protons}, and {Helium} {Ions} (version
  1.2.3). {Available}: http://physics.nist.gov/{Star}}} (\bibinfo {year}
  {2019})\BibitemShut {NoStop}%
\bibitem [{\citenamefont {Moll}(2018)}]{moll_focused_2018}%
  \BibitemOpen
  \bibfield  {author} {\bibinfo {author} {\bibfnamefont {P.~J.}\ \bibnamefont
  {Moll}},\ }\bibfield  {title} {\bibinfo {title} {Focused {Ion} {Beam}
  {Microstructuring} of {Quantum} {Matter}},\ }\href
  {https://doi.org/10.1146/annurev-conmatphys-033117-054021} {\bibfield
  {journal} {\bibinfo  {journal} {Annual Review of Condensed Matter Physics}\
  }\textbf {\bibinfo {volume} {9}},\ \bibinfo {pages} {147} (\bibinfo {year}
  {2018})}\BibitemShut {NoStop}%
\bibitem [{\citenamefont {Iseler}\ \emph {et~al.}(1966)\citenamefont {Iseler},
  \citenamefont {Dawson}, \citenamefont {Mehner},\ and\ \citenamefont
  {Kauffman}}]{iseler_production_1966}%
  \BibitemOpen
  \bibfield  {author} {\bibinfo {author} {\bibfnamefont {G.~W.}\ \bibnamefont
  {Iseler}}, \bibinfo {author} {\bibfnamefont {H.~I.}\ \bibnamefont {Dawson}},
  \bibinfo {author} {\bibfnamefont {A.~S.}\ \bibnamefont {Mehner}},\ and\
  \bibinfo {author} {\bibfnamefont {J.~W.}\ \bibnamefont {Kauffman}},\
  }\bibfield  {title} {\bibinfo {title} {Production {Rates} of {Electrical}
  {Resistivity} in {Copper} and {Aluminum} {Induced} by {Electron}
  {Irradiation}},\ }\href {https://doi.org/10.1103/PhysRev.146.468} {\bibfield
  {journal} {\bibinfo  {journal} {Physical Review}\ }\textbf {\bibinfo {volume}
  {146}},\ \bibinfo {pages} {468} (\bibinfo {year} {1966})}\BibitemShut
  {NoStop}%
\bibitem [{\citenamefont {Legris}\ \emph {et~al.}(1993)\citenamefont {Legris},
  \citenamefont {Rullier-Albenque}, \citenamefont {Radeva},\ and\ \citenamefont
  {Lejay}}]{legris_effects_1993}%
  \BibitemOpen
  \bibfield  {author} {\bibinfo {author} {\bibfnamefont {A.}~\bibnamefont
  {Legris}}, \bibinfo {author} {\bibfnamefont {F.}~\bibnamefont
  {Rullier-Albenque}}, \bibinfo {author} {\bibfnamefont {E.}~\bibnamefont
  {Radeva}},\ and\ \bibinfo {author} {\bibfnamefont {P.}~\bibnamefont
  {Lejay}},\ }\bibfield  {title} {{\selectlanguage {english}\bibinfo {title}
  {Effects of electron irradiation on
  {YBa}$_{\textrm{2}}${Cu}$_{\textrm{3}}${O}$_{\textrm{7-$\delta$}}$
  superconductor}},\ }\href {https://doi.org/10.1051/jp1:1993203} {\bibfield
  {journal} {\bibinfo  {journal} {Journal de Physique I}\ }\textbf {\bibinfo
  {volume} {3}},\ \bibinfo {pages} {1605} (\bibinfo {year} {1993})}\BibitemShut
  {NoStop}%
\bibitem [{\citenamefont {Meyer}\ \emph {et~al.}(2012)\citenamefont {Meyer},
  \citenamefont {Eder}, \citenamefont {Kurasch}, \citenamefont {Skakalova},
  \citenamefont {Kotakoski}, \citenamefont {Park}, \citenamefont {Roth},
  \citenamefont {Chuvilin}, \citenamefont {Eyhusen}, \citenamefont {Benner},
  \citenamefont {Krasheninnikov},\ and\ \citenamefont
  {Kaiser}}]{meyer_accurate_2012}%
  \BibitemOpen
  \bibfield  {author} {\bibinfo {author} {\bibfnamefont {J.~C.}\ \bibnamefont
  {Meyer}}, \bibinfo {author} {\bibfnamefont {F.}~\bibnamefont {Eder}},
  \bibinfo {author} {\bibfnamefont {S.}~\bibnamefont {Kurasch}}, \bibinfo
  {author} {\bibfnamefont {V.}~\bibnamefont {Skakalova}}, \bibinfo {author}
  {\bibfnamefont {J.}~\bibnamefont {Kotakoski}}, \bibinfo {author}
  {\bibfnamefont {H.~J.}\ \bibnamefont {Park}}, \bibinfo {author}
  {\bibfnamefont {S.}~\bibnamefont {Roth}}, \bibinfo {author} {\bibfnamefont
  {A.}~\bibnamefont {Chuvilin}}, \bibinfo {author} {\bibfnamefont
  {S.}~\bibnamefont {Eyhusen}}, \bibinfo {author} {\bibfnamefont
  {G.}~\bibnamefont {Benner}}, \bibinfo {author} {\bibfnamefont {A.~V.}\
  \bibnamefont {Krasheninnikov}},\ and\ \bibinfo {author} {\bibfnamefont
  {U.}~\bibnamefont {Kaiser}},\ }\bibfield  {title} {\bibinfo {title} {Accurate
  {Measurement} of {Electron} {Beam} {Induced} {Displacement} {Cross}
  {Sections} for {Single}-{Layer} {Graphene}},\ }\href
  {https://doi.org/10.1103/PhysRevLett.108.196102} {\bibfield  {journal}
  {\bibinfo  {journal} {Physical Review Letters}\ }\textbf {\bibinfo {volume}
  {108}},\ \bibinfo {pages} {196102} (\bibinfo {year} {2012})}\BibitemShut
  {NoStop}%
\bibitem [{\citenamefont {Mott}(1929)}]{mott_scattering_1929}%
  \BibitemOpen
  \bibfield  {author} {\bibinfo {author} {\bibfnamefont {N.~F.}\ \bibnamefont
  {Mott}},\ }\bibfield  {title} {{\selectlanguage {english}\bibinfo {title}
  {The {Scattering} of {Fast} {Electrons} by {Atomic} {Nuclei}}},\ }\href
  {https://doi.org/10.1098/rspa.1929.0127} {\bibfield  {journal} {\bibinfo
  {journal} {Proceedings of the Royal Society of London A: Mathematical,
  Physical and Engineering Sciences}\ }\textbf {\bibinfo {volume} {124}},\
  \bibinfo {pages} {425} (\bibinfo {year} {1929})}\BibitemShut {NoStop}%
\bibitem [{\citenamefont {Mott}(1932)}]{mott_polarisation_1932}%
  \BibitemOpen
  \bibfield  {author} {\bibinfo {author} {\bibfnamefont {N.~F.}\ \bibnamefont
  {Mott}},\ }\bibfield  {title} {{\selectlanguage {english}\bibinfo {title}
  {The {Polarisation} of {Electrons} by {Double} {Scattering}}},\ }\href
  {https://doi.org/10.1098/rspa.1932.0044} {\bibfield  {journal} {\bibinfo
  {journal} {Proceedings of the Royal Society of London A: Mathematical,
  Physical and Engineering Sciences}\ }\textbf {\bibinfo {volume} {135}},\
  \bibinfo {pages} {429} (\bibinfo {year} {1932})}\BibitemShut {NoStop}%
\bibitem [{\citenamefont {McKinley}\ and\ \citenamefont
  {Feshbach}(1948)}]{mckinley_coulomb_1948}%
  \BibitemOpen
  \bibfield  {author} {\bibinfo {author} {\bibfnamefont {W.~A.}\ \bibnamefont
  {McKinley}}\ and\ \bibinfo {author} {\bibfnamefont {H.}~\bibnamefont
  {Feshbach}},\ }\bibfield  {title} {\bibinfo {title} {The {Coulomb}
  {Scattering} of {Relativistic} {Electrons} by {Nuclei}},\ }\href
  {https://doi.org/10.1103/PhysRev.74.1759} {\bibfield  {journal} {\bibinfo
  {journal} {Physical Review}\ }\textbf {\bibinfo {volume} {74}},\ \bibinfo
  {pages} {1759} (\bibinfo {year} {1948})}\BibitemShut {NoStop}%
\bibitem [{\citenamefont {Lijian}\ \emph {et~al.}(1995)\citenamefont {Lijian},
  \citenamefont {Qing},\ and\ \citenamefont
  {Zhengming}}]{lijian_analytic_1995}%
  \BibitemOpen
  \bibfield  {author} {\bibinfo {author} {\bibfnamefont {T.}~\bibnamefont
  {Lijian}}, \bibinfo {author} {\bibfnamefont {H.}~\bibnamefont {Qing}},\ and\
  \bibinfo {author} {\bibfnamefont {L.}~\bibnamefont {Zhengming}},\ }\bibfield
  {title} {{\selectlanguage {english}\bibinfo {title} {Analytic fitting to the
  mott cross section of electrons}},\ }\href
  {https://doi.org/10.1016/0969-806X(94)00063-8} {\bibfield  {journal}
  {\bibinfo  {journal} {Radiation Physics and Chemistry}\ }\textbf {\bibinfo
  {volume} {45}},\ \bibinfo {pages} {235} (\bibinfo {year} {1995})}\BibitemShut
  {NoStop}%
\bibitem [{\citenamefont {Boschini}\ \emph {et~al.}(2013)\citenamefont
  {Boschini}, \citenamefont {Consolandi}, \citenamefont {Gervasi},
  \citenamefont {Giani}, \citenamefont {Grandi}, \citenamefont {Ivanchenko},
  \citenamefont {Nieminem}, \citenamefont {Pensotti}, \citenamefont
  {Rancoita},\ and\ \citenamefont {Tacconi}}]{boschini_expression_2013}%
  \BibitemOpen
  \bibfield  {author} {\bibinfo {author} {\bibfnamefont {M.~J.}\ \bibnamefont
  {Boschini}}, \bibinfo {author} {\bibfnamefont {C.}~\bibnamefont
  {Consolandi}}, \bibinfo {author} {\bibfnamefont {M.}~\bibnamefont {Gervasi}},
  \bibinfo {author} {\bibfnamefont {S.}~\bibnamefont {Giani}}, \bibinfo
  {author} {\bibfnamefont {D.}~\bibnamefont {Grandi}}, \bibinfo {author}
  {\bibfnamefont {V.}~\bibnamefont {Ivanchenko}}, \bibinfo {author}
  {\bibfnamefont {P.}~\bibnamefont {Nieminem}}, \bibinfo {author}
  {\bibfnamefont {S.}~\bibnamefont {Pensotti}}, \bibinfo {author}
  {\bibfnamefont {P.~G.}\ \bibnamefont {Rancoita}},\ and\ \bibinfo {author}
  {\bibfnamefont {M.}~\bibnamefont {Tacconi}},\ }\bibfield  {title}
  {{\selectlanguage {english}\bibinfo {title} {An expression for the {Mott}
  cross section of electrons and positrons on nuclei with {Z} up to 118}},\
  }\href {https://doi.org/10.1016/j.radphyschem.2013.04.020} {\bibfield
  {journal} {\bibinfo  {journal} {Radiation Physics and Chemistry}\ }\textbf
  {\bibinfo {volume} {90}},\ \bibinfo {pages} {39} (\bibinfo {year}
  {2013})}\BibitemShut {NoStop}%
\bibitem [{Note1()}]{Note1}%
  \BibitemOpen
  \bibinfo {note} {$\beta =\protect \sqrt {1-(m_{e}c^2/(m_{e}c^2+E_{K}))^2}$,
  where $m_{e}$ is the electron rest mass.}\BibitemShut {Stop}%
\bibitem [{Note2()}]{Note2}%
  \BibitemOpen
  \bibinfo {note} {The scattering angle $\vartheta $ and the transferred energy
  $E$ are related by: $E=E_{max}\protect \qopname \relax o{sin}^2{\vartheta
  /2}$. $E_{max}$ is the maximum energy that can be transferred,
  $E_{max}=2E_{K}\left (E_{K}+2m_{e}c^2\right )/Mc^2$. $E_{K}$ is the kinetic
  energy of the incoming electron, and $m_{e}$ and $M$ the masses of the
  electron and the nucleus, respectively.}\BibitemShut {Stop}%
\bibitem [{Note3()}]{Note3}%
  \BibitemOpen
  \bibinfo {note} {Technically the accelerator energy can be reduced down to
  $\protect \unit [0.3]{MeV}$, however this requires a stabilisation time of a
  few days, and was therefore not feasible in the limited beamtime}\BibitemShut
  {NoStop}%
\bibitem [{\citenamefont {Kikugawa}\ and\ \citenamefont
  {Maeno}(2002)}]{kikugawa_non-fermi-liquid_2002}%
  \BibitemOpen
  \bibfield  {author} {\bibinfo {author} {\bibfnamefont {N.}~\bibnamefont
  {Kikugawa}}\ and\ \bibinfo {author} {\bibfnamefont {Y.}~\bibnamefont
  {Maeno}},\ }\bibfield  {title} {\bibinfo {title} {Non-{Fermi}-{Liquid}
  {Behavior} in {Sr}$_{\textrm{2}}${RuO}$_{\textrm{4}}$ with {Nonmagnetic}
  {Impurities}},\ }\href {https://doi.org/10.1103/PhysRevLett.89.117001}
  {\bibfield  {journal} {\bibinfo  {journal} {Physical Review Letters}\
  }\textbf {\bibinfo {volume} {89}},\ \bibinfo {pages} {117001} (\bibinfo
  {year} {2002})}\BibitemShut {NoStop}%
\bibitem [{\citenamefont {Fukuzumi}\ \emph {et~al.}(1996)\citenamefont
  {Fukuzumi}, \citenamefont {Mizuhashi}, \citenamefont {Takenaka},\ and\
  \citenamefont {Uchida}}]{fukuzumi_universal_1996}%
  \BibitemOpen
  \bibfield  {author} {\bibinfo {author} {\bibfnamefont {Y.}~\bibnamefont
  {Fukuzumi}}, \bibinfo {author} {\bibfnamefont {K.}~\bibnamefont {Mizuhashi}},
  \bibinfo {author} {\bibfnamefont {K.}~\bibnamefont {Takenaka}},\ and\
  \bibinfo {author} {\bibfnamefont {S.}~\bibnamefont {Uchida}},\ }\bibfield
  {title} {\bibinfo {title} {Universal {Superconductor}-{Insulator}
  {Transition} and {T}$_{\textrm{c}}$ {Depression} in {Zn}-{Substituted}
  {High}- {T}$_{\textrm{c}}$ {Cuprates} in the {Underdoped} {Regime}},\ }\href
  {https://doi.org/10.1103/PhysRevLett.76.684} {\bibfield  {journal} {\bibinfo
  {journal} {Physical Review Letters}\ }\textbf {\bibinfo {volume} {76}},\
  \bibinfo {pages} {684} (\bibinfo {year} {1996})}\BibitemShut {NoStop}%
\bibitem [{\citenamefont {Rullier-Albenque}\ \emph {et~al.}(2000)\citenamefont
  {Rullier-Albenque}, \citenamefont {Vieillefond}, \citenamefont {Alloul},
  \citenamefont {Tyler}, \citenamefont {Lejay},\ and\ \citenamefont
  {Marucco}}]{rullier-albenque_universal_2000}%
  \BibitemOpen
  \bibfield  {author} {\bibinfo {author} {\bibfnamefont {F.}~\bibnamefont
  {Rullier-Albenque}}, \bibinfo {author} {\bibfnamefont {P.~A.}\ \bibnamefont
  {Vieillefond}}, \bibinfo {author} {\bibfnamefont {H.}~\bibnamefont {Alloul}},
  \bibinfo {author} {\bibfnamefont {A.~W.}\ \bibnamefont {Tyler}}, \bibinfo
  {author} {\bibfnamefont {P.}~\bibnamefont {Lejay}},\ and\ \bibinfo {author}
  {\bibfnamefont {J.~F.}\ \bibnamefont {Marucco}},\ }\bibfield  {title}
  {{\selectlanguage {english}\bibinfo {title} {Universal {T}$_{\textrm{c}}$
  depression by irradiation defects in underdoped and overdoped cuprates?}},\
  }\href {https://doi.org/10.1209/epl/i2000-00238-x} {\bibfield  {journal}
  {\bibinfo  {journal} {EPL}\ }\textbf {\bibinfo {volume} {50}},\ \bibinfo
  {pages} {81} (\bibinfo {year} {2000})}\BibitemShut {NoStop}%
\bibitem [{\citenamefont {Grabowski}\ \emph {et~al.}(2009)\citenamefont
  {Grabowski}, \citenamefont {Ismer}, \citenamefont {Hickel},\ and\
  \citenamefont {Neugebauer}}]{grabowski_ab_2009}%
  \BibitemOpen
  \bibfield  {author} {\bibinfo {author} {\bibfnamefont {B.}~\bibnamefont
  {Grabowski}}, \bibinfo {author} {\bibfnamefont {L.}~\bibnamefont {Ismer}},
  \bibinfo {author} {\bibfnamefont {T.}~\bibnamefont {Hickel}},\ and\ \bibinfo
  {author} {\bibfnamefont {J.}~\bibnamefont {Neugebauer}},\ }\bibfield  {title}
  {\bibinfo {title} {Ab initio up to the melting point: {Anharmonicity} and
  vacancies in aluminum},\ }\href {https://doi.org/10.1103/PhysRevB.79.134106}
  {\bibfield  {journal} {\bibinfo  {journal} {Physical Review B}\ }\textbf
  {\bibinfo {volume} {79}},\ \bibinfo {pages} {134106} (\bibinfo {year}
  {2009})}\BibitemShut {NoStop}%
\bibitem [{\citenamefont {Freysoldt}\ \emph {et~al.}(2014)\citenamefont
  {Freysoldt}, \citenamefont {Grabowski}, \citenamefont {Hickel}, \citenamefont
  {Neugebauer}, \citenamefont {Kresse}, \citenamefont {Janotti},\ and\
  \citenamefont {Van~de Walle}}]{freysoldt_first-principles_2014}%
  \BibitemOpen
  \bibfield  {author} {\bibinfo {author} {\bibfnamefont {C.}~\bibnamefont
  {Freysoldt}}, \bibinfo {author} {\bibfnamefont {B.}~\bibnamefont
  {Grabowski}}, \bibinfo {author} {\bibfnamefont {T.}~\bibnamefont {Hickel}},
  \bibinfo {author} {\bibfnamefont {J.}~\bibnamefont {Neugebauer}}, \bibinfo
  {author} {\bibfnamefont {G.}~\bibnamefont {Kresse}}, \bibinfo {author}
  {\bibfnamefont {A.}~\bibnamefont {Janotti}},\ and\ \bibinfo {author}
  {\bibfnamefont {C.~G.}\ \bibnamefont {Van~de Walle}},\ }\bibfield  {title}
  {\bibinfo {title} {First-principles calculations for point defects in
  solids},\ }\href {https://doi.org/10.1103/RevModPhys.86.253} {\bibfield
  {journal} {\bibinfo  {journal} {Reviews of Modern Physics}\ }\textbf
  {\bibinfo {volume} {86}},\ \bibinfo {pages} {253} (\bibinfo {year}
  {2014})}\BibitemShut {NoStop}%
\bibitem [{\citenamefont {Foiles}\ \emph {et~al.}(1986)\citenamefont {Foiles},
  \citenamefont {Baskes},\ and\ \citenamefont
  {Daw}}]{foiles_embedded-atom-method_1986}%
  \BibitemOpen
  \bibfield  {author} {\bibinfo {author} {\bibfnamefont {S.~M.}\ \bibnamefont
  {Foiles}}, \bibinfo {author} {\bibfnamefont {M.~I.}\ \bibnamefont {Baskes}},\
  and\ \bibinfo {author} {\bibfnamefont {M.~S.}\ \bibnamefont {Daw}},\
  }\bibfield  {title} {\bibinfo {title} {Embedded-atom-method functions for the
  fcc metals {Cu}, {Ag}, {Au}, {Ni}, {Pd}, {Pt}, and their alloys},\ }\href
  {https://doi.org/10.1103/PhysRevB.33.7983} {\bibfield  {journal} {\bibinfo
  {journal} {Phys. Rev. B}\ }\textbf {\bibinfo {volume} {33}},\ \bibinfo
  {pages} {7983} (\bibinfo {year} {1986})}\BibitemShut {NoStop}%
\bibitem [{\citenamefont {Mattsson}\ and\ \citenamefont
  {Mattsson}(2002)}]{mattsson_calculating_2002}%
  \BibitemOpen
  \bibfield  {author} {\bibinfo {author} {\bibfnamefont {T.~R.}\ \bibnamefont
  {Mattsson}}\ and\ \bibinfo {author} {\bibfnamefont {A.~E.}\ \bibnamefont
  {Mattsson}},\ }\bibfield  {title} {\bibinfo {title} {Calculating the vacancy
  formation energy in metals: {Pt}, {Pd}, and {Mo}},\ }\href
  {https://doi.org/10.1103/PhysRevB.66.214110} {\bibfield  {journal} {\bibinfo
  {journal} {Phys. Rev. B}\ }\textbf {\bibinfo {volume} {66}},\ \bibinfo
  {pages} {214110} (\bibinfo {year} {2002})}\BibitemShut {NoStop}%
\bibitem [{\citenamefont {Nazarov}\ \emph {et~al.}(2012)\citenamefont
  {Nazarov}, \citenamefont {Hickel},\ and\ \citenamefont
  {Neugebauer}}]{nazarov_vacancy_2012}%
  \BibitemOpen
  \bibfield  {author} {\bibinfo {author} {\bibfnamefont {R.}~\bibnamefont
  {Nazarov}}, \bibinfo {author} {\bibfnamefont {T.}~\bibnamefont {Hickel}},\
  and\ \bibinfo {author} {\bibfnamefont {J.}~\bibnamefont {Neugebauer}},\
  }\bibfield  {title} {\bibinfo {title} {Vacancy formation energies in fcc
  metals: {Influence} of exchange-correlation functionals and correction
  schemes},\ }\href {https://doi.org/10.1103/PhysRevB.85.144118} {\bibfield
  {journal} {\bibinfo  {journal} {Phys. Rev. B}\ }\textbf {\bibinfo {volume}
  {85}},\ \bibinfo {pages} {144118} (\bibinfo {year} {2012})}\BibitemShut
  {NoStop}%
\bibitem [{\citenamefont {Schaefer}(1987)}]{schaefer_investigation_1987}%
  \BibitemOpen
  \bibfield  {author} {\bibinfo {author} {\bibfnamefont {H.-E.}\ \bibnamefont
  {Schaefer}},\ }\bibfield  {title} {\bibinfo {title} {Investigation of
  {Thermal} {Equilibrium} {Vacancies} in {Metals} by {Positron}
  {Annihilation}},\ }\href {https://doi.org/10.1002/pssa.2211020104} {\bibfield
   {journal} {\bibinfo  {journal} {physica status solidi (a)}\ }\textbf
  {\bibinfo {volume} {102}},\ \bibinfo {pages} {47} (\bibinfo {year}
  {1987})}\BibitemShut {NoStop}%
\bibitem [{\citenamefont {Ullmaier}\ \emph {et~al.}(1991)\citenamefont
  {Ullmaier}, \citenamefont {Ehrhart}, \citenamefont {Jung},\ and\
  \citenamefont {Schultz}}]{ullmaier_atomic_1991}%
  \BibitemOpen
  \bibfield  {author} {\bibinfo {author} {\bibfnamefont {H.}~\bibnamefont
  {Ullmaier}}, \bibinfo {author} {\bibfnamefont {P.}~\bibnamefont {Ehrhart}},
  \bibinfo {author} {\bibfnamefont {P.}~\bibnamefont {Jung}},\ and\ \bibinfo
  {author} {\bibfnamefont {H.}~\bibnamefont {Schultz}},\ }\href@noop {} {\emph
  {\bibinfo {title} {Atomic defects in metals}}},\ Vol.~\bibinfo {volume} {3}\
  (\bibinfo  {publisher} {Springer},\ \bibinfo {year} {1991})\BibitemShut
  {NoStop}%
\bibitem [{\citenamefont {Ziman}(1972)}]{ziman_principles_1972}%
  \BibitemOpen
  \bibfield  {author} {\bibinfo {author} {\bibfnamefont {J.~M.}\ \bibnamefont
  {Ziman}},\ }\href@noop {} {{\selectlanguage {english}\emph {\bibinfo {title}
  {Principles of the {Theory} of {Solids}}}}}\ (\bibinfo  {publisher}
  {Cambridge University Press},\ \bibinfo {year} {1972})\BibitemShut {NoStop}%
\bibitem [{\citenamefont {Harris}\ \emph {et~al.}(1995)\citenamefont {Harris},
  \citenamefont {Yan}, \citenamefont {Matl}, \citenamefont {Ong}, \citenamefont
  {Anderson}, \citenamefont {Kimura},\ and\ \citenamefont
  {Kitazawa}}]{harris_violation_1995}%
  \BibitemOpen
  \bibfield  {author} {\bibinfo {author} {\bibfnamefont {J.~M.}\ \bibnamefont
  {Harris}}, \bibinfo {author} {\bibfnamefont {Y.~F.}\ \bibnamefont {Yan}},
  \bibinfo {author} {\bibfnamefont {P.}~\bibnamefont {Matl}}, \bibinfo {author}
  {\bibfnamefont {N.~P.}\ \bibnamefont {Ong}}, \bibinfo {author} {\bibfnamefont
  {P.~W.}\ \bibnamefont {Anderson}}, \bibinfo {author} {\bibfnamefont
  {T.}~\bibnamefont {Kimura}},\ and\ \bibinfo {author} {\bibfnamefont
  {K.}~\bibnamefont {Kitazawa}},\ }\bibfield  {title} {\bibinfo {title}
  {Violation of {Kohler}'s {Rule} in the {Normal}-{State} {Magnetoresistance}
  of {YBa}$_{\textrm{2}}${Cu}$_{\textrm{3}}${O}$_{\textrm{7-$\delta$}}$ and
  {La}$_{\textrm{2}}$ {Sr}$_{\textrm{x}}${CuO}$_{\textrm{4}}$},\ }\href
  {https://doi.org/10.1103/PhysRevLett.75.1391} {\bibfield  {journal} {\bibinfo
   {journal} {Physical Review Letters}\ }\textbf {\bibinfo {volume} {75}},\
  \bibinfo {pages} {1391} (\bibinfo {year} {1995})}\BibitemShut {NoStop}%
\bibitem [{\citenamefont {Groth}\ \emph {et~al.}(2014)\citenamefont {Groth},
  \citenamefont {Wimmer}, \citenamefont {Akhmerov},\ and\ \citenamefont
  {Waintal}}]{groth_kwant:_2014}%
  \BibitemOpen
  \bibfield  {author} {\bibinfo {author} {\bibfnamefont {C.~W.}\ \bibnamefont
  {Groth}}, \bibinfo {author} {\bibfnamefont {M.}~\bibnamefont {Wimmer}},
  \bibinfo {author} {\bibfnamefont {A.~R.}\ \bibnamefont {Akhmerov}},\ and\
  \bibinfo {author} {\bibfnamefont {X.}~\bibnamefont {Waintal}},\ }\bibfield
  {title} {{\selectlanguage {english}\bibinfo {title} {Kwant: a software
  package for quantum transport}},\ }\href
  {https://doi.org/10.1088/1367-2630/16/6/063065} {\bibfield  {journal}
  {\bibinfo  {journal} {New Journal of Physics}\ }\textbf {\bibinfo {volume}
  {16}},\ \bibinfo {pages} {063065} (\bibinfo {year} {2014})}\BibitemShut
  {NoStop}%
\bibitem [{\citenamefont {Garcia}\ \emph {et~al.}(2019)\citenamefont {Garcia},
  \citenamefont {Coulter},\ and\ \citenamefont
  {Narang}}]{garcia_optoelectronic_2019}%
  \BibitemOpen
  \bibfield  {author} {\bibinfo {author} {\bibfnamefont {C.~A.~C.}\
  \bibnamefont {Garcia}}, \bibinfo {author} {\bibfnamefont {J.}~\bibnamefont
  {Coulter}},\ and\ \bibinfo {author} {\bibfnamefont {P.}~\bibnamefont
  {Narang}},\ }\bibfield  {title} {\bibinfo {title} {Optoelectronic {Response}
  of {Type}-{I} {Weyl} {Semimetals} {TaAs} and {NbAs} from {First}
  {Principles}},\ }\href {http://arxiv.org/abs/1907.04348} {\bibfield
  {journal} {\bibinfo  {journal} {arXiv:1907.04348}\ } (\bibinfo {year}
  {2019})}\BibitemShut {NoStop}%
\bibitem [{\citenamefont {Coleridge}(1991)}]{coleridge_small-angle_1991}%
  \BibitemOpen
  \bibfield  {author} {\bibinfo {author} {\bibfnamefont {P.~T.}\ \bibnamefont
  {Coleridge}},\ }\bibfield  {title} {\bibinfo {title} {Small-angle scattering
  in two-dimensional electron gases},\ }\href
  {https://doi.org/10.1103/PhysRevB.44.3793} {\bibfield  {journal} {\bibinfo
  {journal} {Physical Review B}\ }\textbf {\bibinfo {volume} {44}},\ \bibinfo
  {pages} {3793} (\bibinfo {year} {1991})}\BibitemShut {NoStop}%
\bibitem [{\citenamefont {S{\'o}lyom}(2008)}]{solyom_fundamentals_2008}%
  \BibitemOpen
  \bibfield  {author} {\bibinfo {author} {\bibfnamefont {J.}~\bibnamefont
  {S{\'o}lyom}},\ }\href@noop {} {{\selectlanguage {English}\emph {\bibinfo
  {title} {Fundamentals of the {Physics} of {Solids}: {Volume} {II}:
  {Electronic} {Properties}}}}}\ (\bibinfo  {publisher} {Springer},\ \bibinfo
  {address} {Berlin},\ \bibinfo {year} {2008})\BibitemShut {NoStop}%
\bibitem [{\citenamefont {Lapidus}(1982)}]{lapidus_quantummechanical_1982}%
  \BibitemOpen
  \bibfield  {author} {\bibinfo {author} {\bibfnamefont {I.~R.}\ \bibnamefont
  {Lapidus}},\ }\bibfield  {title} {\bibinfo {title} {Quantum-mechanical
  scattering in two dimensions},\ }\href {https://doi.org/10.1119/1.13004}
  {\bibfield  {journal} {\bibinfo  {journal} {American Journal of Physics}\
  }\textbf {\bibinfo {volume} {50}},\ \bibinfo {pages} {45} (\bibinfo {year}
  {1982})}\BibitemShut {NoStop}%
\bibitem [{\citenamefont {Perdew}\ \emph {et~al.}(1996)\citenamefont {Perdew},
  \citenamefont {Burke},\ and\ \citenamefont
  {Ernzerhof}}]{perdew_generalized_1996}%
  \BibitemOpen
  \bibfield  {author} {\bibinfo {author} {\bibfnamefont {J.~P.}\ \bibnamefont
  {Perdew}}, \bibinfo {author} {\bibfnamefont {K.}~\bibnamefont {Burke}},\ and\
  \bibinfo {author} {\bibfnamefont {M.}~\bibnamefont {Ernzerhof}},\ }\bibfield
  {title} {\bibinfo {title} {Generalized {Gradient} {Approximation} {Made}
  {Simple}},\ }\href {https://doi.org/10.1103/PhysRevLett.77.3865} {\bibfield
  {journal} {\bibinfo  {journal} {Physical Review Letters}\ }\textbf {\bibinfo
  {volume} {77}},\ \bibinfo {pages} {3865} (\bibinfo {year}
  {1996})}\BibitemShut {NoStop}%
\bibitem [{\citenamefont {Kresse}\ and\ \citenamefont
  {Furthm{\"u}ller}(1996)}]{kresse_efficient_1996}%
  \BibitemOpen
  \bibfield  {author} {\bibinfo {author} {\bibfnamefont {G.}~\bibnamefont
  {Kresse}}\ and\ \bibinfo {author} {\bibfnamefont {J.}~\bibnamefont
  {Furthm{\"u}ller}},\ }\bibfield  {title} {\bibinfo {title} {Efficient
  iterative schemes for ab initio total-energy calculations using a plane-wave
  basis set},\ }\href {https://doi.org/10.1103/PhysRevB.54.11169} {\bibfield
  {journal} {\bibinfo  {journal} {Physical Review B}\ }\textbf {\bibinfo
  {volume} {54}},\ \bibinfo {pages} {11169} (\bibinfo {year}
  {1996})}\BibitemShut {NoStop}%
\bibitem [{\citenamefont {Ong}\ \emph {et~al.}(2010)\citenamefont {Ong},
  \citenamefont {Zhang}, \citenamefont {Tse},\ and\ \citenamefont
  {Wu}}]{ong_origin_2010}%
  \BibitemOpen
  \bibfield  {author} {\bibinfo {author} {\bibfnamefont {K.~P.}\ \bibnamefont
  {Ong}}, \bibinfo {author} {\bibfnamefont {J.}~\bibnamefont {Zhang}}, \bibinfo
  {author} {\bibfnamefont {J.~S.}\ \bibnamefont {Tse}},\ and\ \bibinfo {author}
  {\bibfnamefont {P.}~\bibnamefont {Wu}},\ }\bibfield  {title} {\bibinfo
  {title} {Origin of anisotropy and metallic behavior in delafossite
  {PdCoO}$_{\textrm{2}}$},\ }\href {https://doi.org/10.1103/PhysRevB.81.115120}
  {\bibfield  {journal} {\bibinfo  {journal} {Physical Review B}\ }\textbf
  {\bibinfo {volume} {81}},\ \bibinfo {pages} {115120} (\bibinfo {year}
  {2010})}\BibitemShut {NoStop}%
\end{thebibliography}%

\end{document}